\documentclass[onecolumn,showpacs,prc,superscriptaddress,amsmath,amssymb,preprintnumbers]{revtex4-1}

\newcommand{ \be }{\begin{eqnarray}}
\newcommand{ \ee }{\end{eqnarray}}

\usepackage{dcolumn} 
\usepackage{bm} 

\usepackage{graphicx}
\usepackage{color}
\definecolor{dgreen}{cmyk}{1.,0.,1.,0.4}        
\definecolor{orange}{cmyk}{0.,0.353,1.,0.}    

\usepackage[pdftex,bookmarks]{hyperref}

\usepackage{lineno}

\begin{document}


\title{Higher order moments of multiparticle azimuthal correlations}
\author{Ante Bilandzic} 
\affiliation{Niels Bohr Institute, University of Copenhagen, Blegdamsvej 17, 2100 Copenhagen, Denmark}
\date{\today}

\begin{abstract}
We introduce a general procedure to pave the road towards the ultimate goal of deriving analytic expressions for the probability density functions (p.d.f.'s) of multiparticle azimuthal correlations. All multiparticle azimuthal correlators can be expressed analytically in terms of the real and imaginary parts of $M$-particle $Q$-vectors. We derive the analytic results for the p.d.f.'s of single-particle $Q$-vectors in the most general case and demonstrate that they can be expressed solely in terms of Chebyshev polynomials of the first kind. This leads analytically to the expressions of the characteristic functions of $M$-particle $Q$-vectors in terms of Bessel functions of the first kind. From the obtained characteristics functions we calculate the higher order moments of the real and imaginary parts of $M$-particle $Q$-vectors and use them to obtain the higher order moments of multiparticle azimuthal correlators. Finally, these results are used to investigate the sensitivity of multiparticle azimuthal correlations and to illuminate requirements necessary for future anisotropic flow measurements.
\end{abstract}

\pacs{25.75.Ld, 25.75.Gz, 05.70.Fh}


\maketitle







\section{Introduction}
\label{s:Introduction}

Multiparticle azimuthal correlations are nowadays utilized regularly in anisotropic flow analysis by all major collaborations worldwide which are analyzing heavy-ion data. In~\cite{Borghini:2000sa} they were introduced in their present form into the experimental heavy-ion physics as an improvement over the standard Event Plane method~\cite{Poskanzer:1998yz} and the two-particle correlation techniques~\cite{Wang:1991qh} after if was realized that both of them are severely biased by contributions from unwanted sources of correlations typically involving only a few particles. Such contributions in general are not related to correlations which originate from the collective nature of anisotropic flow, which typically involve all produced particles. A great deal of effort has been invested in the last fifteen years in developing the technology to enable efficient and exact evaluation of multiparticle azimuthal correlations, which would be free from trivial (yet dominant) contributions from autocorrelations. For this reason the sophisticated and efficient formalism of generating functions was  proposed in~\cite{Borghini:2000sa,Borghini:2001vi}. However, despite its initial success it was eventually realized that this formalism is non-exact and limited in scope only to certain types of multiparticle correlations. Recently the generic framework was developed in~\cite{Bilandzic:2013kga} which enables exact and efficient evaluation of any multiparticle azimuthal correlation (for a detailed historical account on the development and usage of multiparticle azimuthal correlations in anisotropic flow analysis we refer to the introductory part of~\cite{Bilandzic:2013kga}).

Although widely used, very little is known about the statistical properties of multiparticle azimuthal correlations beyond their first moments. In this technical paper we therefore initiate the work which attempts to fill in this gap in the literature---the derivation of higher order moments and the derivation of the analytic expressions for the probability density functions (p.d.f.'s) of multiparticle azimuthal correlations. If the anisotropic flow is the only source of correlations among produced particles and is quantified in the standard way, namely in terms of flow harmonics $v_n$ and symmetry planes $\Psi_n$ which appear as two distinct degrees of freedom in a Fourier series expansion of anisotropic distribution of produced particles in momentum space~\cite{Voloshin:1994mz}, then the following analytic expression, derived first in~\cite{Bhalerao:2011yg}, follows for the first moments of multiparticle azimuthal correlations:
\begin{equation}
\mu_{\left<m\right>_{n_1,n_2,\ldots,n_m}} = v_{n_1}\cdots v_{n_m}e^{i(n_1\Psi_{n_{1}}+\cdots+n_m\Psi_{n_m})}\,.
\label{eq:mixedHarmonicsExpVersion}
\end{equation}
In the above relation by $\left<m\right>_{n_1,n_2,\ldots,n_m}$ we have denoted the generic $m$-particle azimuthal correlation evaluated in harmonics $n_1,n_2,\ldots,n_m$ (for a detailed definition and further discussion see~\cite{Bilandzic:2013kga}). The key point in the derivation of result (\ref{eq:mixedHarmonicsExpVersion}) was the assumption that the collective anisotropic flow is the only source of correlations between produced particles. This assumption implies that a joint $n$-variate p.d.f., $f(\varphi_1,\ldots,\varphi_n)$, will factorize into the product of $n$ single-particle marginalized p.d.f.'s  $f_{\varphi_i}(\varphi_i)$, $1\leq i\leq n$:
\begin{equation}
f(\varphi_1,\ldots,\varphi_n) = f_{\varphi_{1}}(\varphi_{1})\cdots f_{\varphi_{n}}(\varphi_n)\,,
\label{eq:factorization}
\end{equation}
where $\varphi$ labels the azimuthal angles of produced particles. In this work we will always assume that the factorization (\ref{eq:factorization}) is exactly satisfied and that the functional form of each individual p.d.f. $f_{\varphi_i}(\varphi_i)$ is exactly the same~\cite{Danielewicz:1983we} and is given by the Fourier series~\cite{Voloshin:1994mz} (for this reason, we will frequently use the terminology {\it Fourier-like} p.d.f. in this paper). 

Of particular interest are isotropic multiparticle azimuthal correlations, for which $n_1+n_2+\cdots+n_m = 0$~\cite{Bhalerao:2011yg}, evaluated in a single harmonic $n$  (i.e. in addition $|n_i|=n$ holds for any $i$). For such specific correlations any dependence on symmetry planes $\Psi_{n_i}$ cancels out in (\ref{eq:mixedHarmonicsExpVersion}) and therefore one can use the first algebraic moment of $m$-particle azimuthal correlation to estimate directly the $m$th algebraic moment of the flow harmonic $v_n$ (to be denoted by $E[v_n^m]$ or $\left<v_n^m\right>$). 
These higher order moments $\left<v_n^m\right>$ are particularly important in the study of physical processes which govern event-by-event flow fluctuations, when each higher order moment by definition provides an independent information on the underlying p.d.f. of flow fluctuations. For a recent theoretical review on flow fluctuations we refer to~\cite{Luzum:2013yya}, while for a recent review of the experimental results on flow fluctuations we refer to~\cite{Snellings:2014kwa,Jia:2014jca}.
Finally, we remark that none of the first moments of isotropic single-harmonic multiparticle azimuthal correlations can be used to estimate directly the very first moment of $v_n$, i.e. its mean value $\left<v_n\right>$. The mean value for instance can be estimated directly in a completely different approach which does not rely on correlation techniques and was published recently in~\cite{Naselsky:2012nw}, or via the unfolding methods~\cite{Jia:2013tja,Aad:2013xma}.

In general, the factorization  (\ref{eq:factorization}) will break down due to the presence of correlations which are involving only a few particles and all of which are commonly referred to as {\it nonflow}. In such a case multiparticle azimuthal correlations will be systematically biased by contributions from nonflow and are not reliable estimators of the anisotropic flow properties. The further improvement in this context came with multiparticle cumulants. In the way they were originally deployed into the anisotropic flow analysis~\cite{Borghini:2000sa,Borghini:2001vi}, the higher order multiparticle cumulants were intended to be systematically less sensitive to unwanted nonflow correlations than the lower order ones. In this work we assume that nonflow is absent and that the factorization (\ref{eq:factorization}) holds exactly, while the systematical biases in multiparticle azimuthal correlations due to nonflow will be addressed in our subsequent work. 

The paper is organized as follows. In Section~\ref{s:On_the_road_towards_the_p.d.f.} we motivate our work, define all observables of interest, present and discuss our main results and indicate the future directions of this project. Our main results are the analytic expressions for the p.d.f.'s of the real and imaginary parts of single-particle $Q$-vectors and the analytic expressions for the characteristic functions of the real and imaginary parts of $M$-particle $Q$-vectors, where $M$ is the total number of particles in an event. These results were obtained for the most general case of multichromatic anisotropic flow and without invoking the central limit theorem. The former results can be expressed solely in terms of Chebyshev polynomials of the first kind, while the latter ones can be expressed solely in terms of Bessel functions of the first kind. In Section~\ref{s:Higher_order_moments} we present the analytic results for the higher order moments of the real and imaginary parts of $M$-particle $Q$-vectors, and utilize them to calculate higher order moments of isotropic multiparticle azimuthal correlations of interest. These results are used in the discussion on sensitivity of correlation techniques which shall be useful in the design considerations of the future detectors aiming at anisotropic flow measurements with multiparticle azimuthal correlations. In each appendix we provide self-contained materials with all technical steps detailed which were omitted in the derivation of the results presented in the main part of paper.   


\section{On the road towards the p.d.f.}
\label{s:On_the_road_towards_the_p.d.f.}

One of the most important observables in the experimental anisotropic flow analysis is the $Q$-vector (sometimes also called the flow vector)~\cite{Ollitrault:1992bk,Barrette:1994xr,Voloshin:1994mz}. The $Q$-vector evaluated in harmonic $n$ is a complex number which is defined for a set of $M$ particles as: 
\begin{equation}
Q_{n} \equiv \sum_{k=1}^{M}\,e^{in\varphi_k} \,,
\label{eq:Qvector}
\end{equation}
where $\varphi_k$ labels the azimuthal angle of $k$th particle. In this work we will make a notable difference between the {\it $M$-particle} $Q$-vector defined above, and {\it single-particle} or {\it unit} $Q$-vector evaluated in harmonic $n$, which is denoted by $u_n$ and defined as: 
\begin{equation}
u_{n} \equiv e^{in\varphi} \,.
\label{eq:QvectorSingleParticleOrUnit}
\end{equation}
Depending on the context, the $M$-particle or the single-particle $Q$-vector can be the more suitable observable to work with. The physical interpretation of $M$-particle $Q$-vector is the following---if a set of $M$ azimuthal angles can define a mean or preferred direction in azimuth, than that direction cannot be estimated na\"{i}vely by a sample mean $\sum_{i=1}^M\,\varphi_i/M$, but is instead estimated with the direction in which $M$-particle $Q$-vector points to~\cite{Fisher:1995}. For completeness sake, we also outline the definition of $q_n$, which is called the {\it reduced} $Q$-vector and defined as:
\begin{equation}
q_n\equiv\frac{Q_n}{\sqrt{M}}\,.
\label{eq:QvectorReduced}
\end{equation}
The reduced $Q$-vector is less sensitive to the biases originating from multiplicity fluctuations and is a particularly suitable observable for the event-shape engineering, which is a newly emerging field of research in heavy-ion collisions gaining a lot of attention of late~\cite{Schukraft:2012ah,Dobrin:2012zx,Petersen:2013vca,Huo:2013qma}. 

In the present work we will utilize $Q$-vectors because of the following remarkable property: All multiparticle azimuthal correlations can be expressed analytically in terms of $M$-particle $Q$-vectors evaluated (in general) in different harmonics~\cite{SV:privateCommunication}. This realization was a major recent breakthrough which eventually led to the exact and efficient evaluation of all multiparticle azimuthal correlations. The key point is that only a single pass over the data suffices to calculate the $Q$-vectors defined in (\ref{eq:Qvector}) for in principle any number of different harmonics $n$. Therefore, the evaluation of analytic expressions which relate multiparticle azimuthal correlations with $Q$-vectors require as well only a single pass over the data. Such analytic equations were published for a few selected and most widely used azimuthal correlations in~\cite{Bilandzic:2010jr}, while the general answer was provided recently in~\cite{Bilandzic:2013kga}. Taking everything into account, we conclude that in order to narrow down the statistical properties of multiparticle azimuthal correlations we are naturally led in the first step towards the study of statistical properties of their fundamental building blocks, the $Q$-vectors. This is the main subject of the current paper. 

Approximate p.d.f.'s, usually derived under the assumption of central limit theorem and under the assumption that only one anisotropic flow harmonic is non-negligible, for various observables which are sensitive to anisotropic flow are already available in the literature. For instance, it was shown in~\cite{Voloshin:1994mz,Voloshin:2007pc} that the Bessel-Gaussian p.d.f. to leading order describes distribution of the modulus of reduced $Q$-vector defined in (\ref{eq:QvectorReduced}). Recently, power-law and elliptic-power p.d.f.'s were studied in~\cite{Yan:2013laa,Yan:2014nsa,Yan:2014afa} which describe the distributions of eccentricities calculated from initial anisotropies in coordinate space. For the calculation of moments of initial anisotropies in the Glauber approach we refer to~\cite{Broniowski:2007ft}, the characterization
of initial conditions with Bessel-Fourier expansion can be found in~\cite{ColemanSmith:2012ka,Floerchinger:2013vua}, while for an alternative characterization with cumulant expansion we refer to~\cite{Teaney:2010vd}. For an overview of experimental results in this context, we refer to the recent review~\cite{Jia:2014jca}. 

In this paper we take an alternative route. Our starting random observable is the azimuthal angle $\varphi$ whose sample space is the interval $[0,2\pi)$ and whose p.d.f. is given by:
\begin{equation}
f(\varphi)=\frac{1}{2\pi}\big[1+2\sum_{n=1}^\infty (c_n\cos n\varphi + s_n\sin n\varphi)\big]\,.
\label{eq:Fourier_cn_sn}
\end{equation}
We have found it more convenient in the following calculations to parametrize Fourier-like p.d.f. with $c_n$ and $s_n$ harmonics instead of with amplitudes $v_n$ and symmetry planes (phases) $\Psi_n$, as has become customary in the field. The parametrization can be trivially changed by using relations $v_n=\sqrt{c_n^2+s_n^2}$ and $\Psi_n=\arctan\frac{s_n}{c_n}$. Any multiparticle azimuthal correlation can be considered as a multivariate function of $M$ azimuthal angles $\varphi_1,\varphi_2,\ldots,\varphi_M$, where $M$ is the total number of particles in an event. Under the assumptions outlined in the Introduction, the joint multivariate p.d.f. of $M$ azimuthal angles $\varphi$ factorizes into the product of $M$ single particle p.d.f.'s, the functional form each of which is given by the relation (\ref{eq:Fourier_cn_sn}) above. The problem of finding p.d.f. of multiparticle azimuthal correlation is therefore at its core equivalent to the problem of finding p.d.f. $g(a)$ of a function $a(x)$ of a starting random observable $x$ whose sample space $X$ and the p.d.f. $f(x)$ are already known. From the conservation of probability it follows~\cite{Cowan:1998ji}:
\begin{equation}
g(a)da = f(x)dx\,.
\label{eq:ConservationOfProbability_MainPart}
\end{equation}
The above result, however, cannot be applied directly in the cases which were encountered in this project. The reason for a limited direct applicability of result (\ref{eq:ConservationOfProbability_MainPart}) lies in the fact that modifications and generalizations are required for the cases when both $a(x)$ and $f(x)$ are piecewise-defined functions, as well as for the case when $a(x)$ doesn't have a unique inverse. We have therefore, in Appendix~\ref{app:Distributions_of_functions_of_random_observables}, provided self-contained material which generalizes the result (\ref{eq:ConservationOfProbability_MainPart}) and we have illustrated in a detailed toy Monte Carlo example how this generalization can be utilized in practice. 

By using the general procedure established in Appendix~\ref{app:Distributions_of_functions_of_random_observables} for finding the p.d.f. of a function of random observable with known p.d.f. we have derived all of our results, which we now present. We consider the real and imaginary parts of single-particle $Q$-vectors evaluated in harmonic $m$ (see Eq.~(\ref{eq:QvectorSingleParticleOrUnit})) as functions of azimuthal angle $\varphi$ whose p.d.f. is given by Eq.~(\ref{eq:Fourier_cn_sn}), and we derive the analytic results for the p.d.f.'s for the real and imaginary parts of single-particle $Q$-vectors for the following cases of interest: a) random walk; b) monochromatic flow; and c) multichromatic flow. By a random walk we mean the case when all harmonics $c_n$ and $s_n$ are zero in a Fourier-like p.d.f. (\ref{eq:Fourier_cn_sn}). On the other hand, by monochromatic flow we mean the case when only one out of all possible harmonics $c_n$ and $s_n$ is non-zero. Finally, by multichromatic flow we mean the most general case when all harmonics $c_n$ and $s_n$ are non-zero. The detailed derivations are provided in Appendix~\ref{app:Fundamental_results}, in the next section we only highlight and discuss the final results. 


\subsection{Random walk}
\label{ss:Random_walk}

For the case when particles are sampled randomly, we have obtained exactly the same results for the p.d.f.'s of the real and imaginary parts of single particle $Q$-vectors. In this case the starting setup is defined as
\begin{equation}
x\in [0,2\pi),\qquad f(x)=\frac{1}{2\pi}, \qquad a(x)=\cos mx {\rm\ or\ } a(x)=\sin mx\,,
\end{equation}
and we have obtained the following analytic result for the p.d.f. $g(a)$:
\begin{equation}
g(a)=\frac{1}{\pi}\frac{1}{\sqrt{1-a^2}}\,,\qquad a\in[-1,1)\,.
\label{eq:RW_main_part}
\end{equation}
We see that the p.d.f. $g(a)$ of both real and imaginary parts of single-particle $Q$-vector evaluated in harmonic $m$ does not depend on the harmonic $m$. The detailed derivation can be found in Appendix~\ref{app:Fundamental_results}.


\subsection{Monochromatic flow}
\label{ss:Monochromatic_flow}

For the case of monochromatic flow when the Fourier-like p.d.f. (\ref{eq:Fourier_cn_sn}) is parametrized with only a single harmonic, either with $c_n$ or $s_n$, there are four distinct cases to be considered. We outline for each case its analytic solution separately---the technical details in the derivation of each case can be found in Appendix~\ref{app:Fundamental_results}, where also a toy Monte Carlo studies were performed independently for each solution.   

We start with the first special case in which we seek the p.d.f. of the real part of single-particle $Q$-vector evaluated in harmonic $m$ and when the starting Fourier-like p.d.f. (\ref{eq:Fourier_cn_sn}) is parametrized only with one harmonic $c_n$ associated with cosine terms. The starting setup is:
\begin{equation}
x\in [0,2\pi),\qquad f(x)=\frac{1}{2\pi}(1+2\,c_n\cos nx),\qquad a(x)=\cos mx\,,
\end{equation}
and we have obtained the following analytic solution for the p.d.f. $g(a)$:
\begin{equation}
g(a)=
\left\{
 \begin{array}{ll}
  \frac{1+2c_nT_{\frac{n}{m}}(a)}{\pi\sqrt{1-a^2}}\,,& n/m{\rm\ is\ arbitrary\ integer}\,, \\
  \frac{1}{\pi\sqrt{1-a^2}}\,,& {\rm otherwise}\,, \\  
 \end{array} 
\right.
\label{eq:g(a)_Cos_Cos_main_part}
\end{equation}
where $a\in[-1,1)$ and $T_n$ is the $n$th Chebyshev polynomial of the first kind. The above solution also applies to negative integers $m$ after trivial replacement of $n/m$ with $n/|m|$ in the index of Chebyshev polynomial.

The second special case is the one in which the p.d.f. of an imaginary part of single-particle $Q$-vector evaluated in harmonic $m$ is obtained when the starting Fourier-like p.d.f. (\ref{eq:Fourier_cn_sn}) is parametrized only with harmonic $c_n$. The starting setup in this case is:
\begin{equation}
x\in [0,2\pi),\qquad f(x)=\frac{1}{2\pi}(1+2c_n\cos nx),\qquad a(x)=\sin mx.
\end{equation}
For the p.d.f. $g(a)$ of the imaginary part of single-particle $Q$-vector we have obtained the following analytic solution:
\begin{equation}
g(a)=
\left\{
 \begin{array}{ll}
  \frac{1+2c_ni^{\frac{n}{m}}T_{\frac{n}{m}}(a)}{\pi\sqrt{1-a^2}}\,,& n/m\ {\rm is\ even\ integer}\,, \\
  \frac{1}{\pi\sqrt{1-a^2}}\,,& {\rm otherwise}\,, \\  
 \end{array} 
\right.
\label{eq:g(a)_Cos_Sin_main_part} 
\end{equation}
where $a\in[-1,1)$ and $T_n$ is the $n$th Chebyshev polynomial of the first kind. The above result can be trivially extended for negative integers $m$ as well, after replacing $m$ with $|m|$ everywhere in Eq.~(\ref{eq:g(a)_Cos_Sin_main_part}) (see Appendix~\ref{app:Fundamental_results} for a detailed discussion which justifies this claim). We also remark on the important role of factor $i^{n/m}$ in (\ref{eq:g(a)_Cos_Sin_main_part}), which evaluates to 1 or -1 depending on whether an even integer $n/m$ in addition satisfies also the relation $n/m\ {\rm mod\ 4} = 0$ (when it evaluates to 1) or not (when it evaluates to -1).

The third distinct case is the simplest one to consider, and it is the case in which the p.d.f. of the real part of single-particle $Q$-vector evaluated in harmonics $m$ is sought when the starting Fourier-like p.d.f. (\ref{eq:Fourier_cn_sn}) is parametrized only with harmonic $s_n$ which is associated with sinus terms. The starting setup for this case is:
\begin{equation}
x\in [0,2\pi),\qquad f(x)=\frac{1}{2\pi}(1+2s_n\sin nx),\qquad a(x)=\cos mx.
\end{equation}
For the p.d.f. $g(a)$ of a real part of single-particle $Q$-vector we have obtained the following analytic solution: 
\begin{equation}
g(a)=\frac{1}{\pi\sqrt{1-a^2}}\,,
\label{eq:g(a)_Sin_Cos_main_part}
\end{equation}
where $a\in [-1,1)$. We see that in this case the final result for any choice of integers $n$ and $m$ is exactly the same as for the random walk (see Eq.~(\ref{eq:RW_main_part})). 

Finally, the last distinct case to consider is the one in which the p.d.f. of an imaginary part of single-particle $Q$-vector is sought when the starting Fourier-like p.d.f. (\ref{eq:Fourier_cn_sn}) is parametrized only with harmonic $s_n$. We start with:
\begin{equation}
x\in [0,2\pi),\qquad f(x)=\frac{1}{2\pi}(1+2s_n\sin nx),\qquad a(x)=\sin mx\,,
\end{equation}
and for the p.d.f. $g(a)$ we have obtained the following analytic solution:
\begin{equation}
g(a)=
\left\{
 \begin{array}{ll}
  \frac{1+2s_n\,i^{\frac{n}{m}-1}T_{\frac{n}{m}}(a)}{\pi\sqrt{1-a^2}}\,,& n/m\ {\rm is\ odd\ integer}\,, \\
  \frac{1}{\pi\sqrt{1-a^2}}\,,& {\rm otherwise}\,, \\  
 \end{array} 
\right.
\label{eq:g(a)_Sin_Sin_main_part}
\end{equation}
where $a\in[-1,1)$ and $T_n$ is the $n$th Chebyshev polynomial of the first kind. With the replacement of $i^{\frac{n}{m}-1}T_{\frac{n}{m}}$ with ${\rm sgn}(m)i^{\frac{n}{|m|}-1}T_{\frac{n}{|m|}}$ in (\ref{eq:g(a)_Sin_Sin_main_part}) this result can be extended also to negative integers $m$ (see Appendix~\ref{app:Fundamental_results} for a detailed further discussion). 

The five results for the p.d.f.'s of the real and imaginary parts of single-particle $Q$-vectors obtained for the random walk, in Eq.~(\ref{eq:RW_main_part}), and for the four distinct cases of monochromatic flow, in Eqs.~(\ref{eq:g(a)_Cos_Cos_main_part}), (\ref{eq:g(a)_Cos_Sin_main_part}), (\ref{eq:g(a)_Sin_Cos_main_part}) and (\ref{eq:g(a)_Sin_Sin_main_part}), are fundamental results and any more general case can be obtained straightforwardly by superimposing these five fundamental results. In the next section we further generalize our results and present them for the most general case of multichromatic flow.  


\subsection{Multichromatic flow}
\label{ss:Multichromatic_flow}

\begin{figure}[h]
\centering
\includegraphics[width=0.5\textwidth]{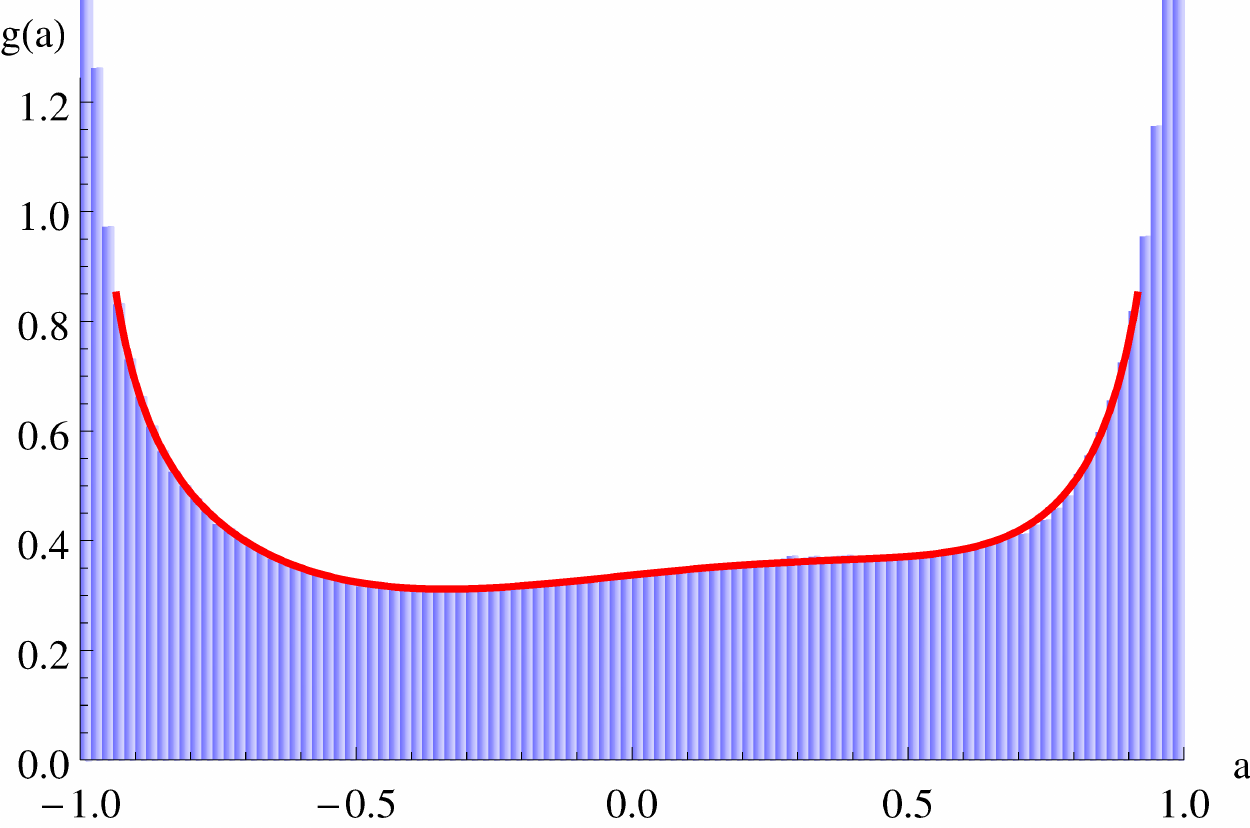}
\caption{(Color online) Distribution of the real part of single-particle $Q$-vector (blue) and its theoretical p.d.f. (solid red line) obtained from Eq.~(\ref{eq:g(a)_fullFS_cosmx_secondVersion_main_part}) in a toy Monte Carlo example for the general case of multichromatic flow.}
\label{fig:generalCombinedCase_1d}
\end{figure}

Having obtained in the previous section all fundamental results, i.e. the results for a random walk and for the four distinct cases of monochromatic flow, the results for the case of multichromatic flow are given trivially as appropriate and straightforward superpositions. In particular, the p.d.f. of the real part of single-particle $Q$-vector for the most general case of multichromatic flow is given as a superposition of the fundamental solutions (\ref{eq:g(a)_Cos_Cos_main_part}) and (\ref{eq:g(a)_Sin_Cos_main_part}). For completeness sake, we outline the starting setup which is defined now as:
\begin{equation}
x\in [0,2\pi),\qquad \displaystyle f(x)=\frac{1}{2\pi}\big[1+2\sum_{n=1}^{\infty}(c_n\cos nx+s_n\sin nx)\big],\qquad a(x)=\cos mx\,.
\end{equation}
For the p.d.f. $g(a)$ of the real part of single-particle $Q$-vector evaluated in harmonic $m$ in this most general case we have obtained the following analytic solution:
\begin{equation}
g(a)=\frac{1}{\pi\sqrt{1-a^2}}\bigg(1+\displaystyle\sum_{\begin{subarray}{c}n\\(n\,{\rm mod}\,m = 0 )\end{subarray}}^{\infty}\,2c_nT_{\frac{n}{m}}(a)\bigg)\,,
\label{eq:g(a)_fullFS_cosmx_firstVersion_main_part}
\end{equation}
or written equivalently:
\begin{equation}
g(a)=\frac{1}{\pi\sqrt{1-a^2}}\left(1+\displaystyle\sum_{l=1}^{\infty}\,2c_{l\cdot m}T_{l}(a)\right)\,.
\label{eq:g(a)_fullFS_cosmx_secondVersion_main_part}
\end{equation}
The above analytic solution (\ref{eq:g(a)_fullFS_cosmx_secondVersion_main_part}) was tested in a toy Monte Carlo example in which we have randomly selected the values for ten harmonics $c_1,c_2,\ldots,c_{10}$ and for ten harmonics $s_1,s_2,\ldots,s_{10}$ from the interval $[0,0.1)$. With those 20 harmonics we have parametrized the Fourier-like p.d.f. in (\ref{eq:Fourier_cn_sn}) and sampled the azimuthal angles distributed according to $f(\varphi)$. From the sampled azimuthal angles we have calculated the real part of the single-particle $Q$-vector, and created its distribution. We remark that not each random selection of harmonics $c_n$ and $s_n$ will result in a physical Fourier like p.d.f. (\ref{eq:Fourier_cn_sn})---the additional requirement on the choice of harmonics is that the resulting p.d.f. $f(x)$ is a positive definite function over the whole interval $[0,2\pi)$, due to its probabilistic nature. For harmonic $m$ we have selected $m=2$, so that $a(x) = \cos 2x$ in this toy Monte Carlo example. On Fig.~\ref{fig:generalCombinedCase_1d} the resulting distribution of $a(x)$ is shown in blue, while the theoretical result for the p.d.f. $g(a)$ calculated from Eq.~(\ref{eq:g(a)_fullFS_cosmx_secondVersion_main_part}) is shown with a solid red line.
Given the result (\ref{eq:g(a)_fullFS_cosmx_secondVersion_main_part}) above, we remark that the distribution of the real part of single-particle $Q$-vector is not sensitive to the presence of harmonics $s_n$ which are associated with the sinus terms in the starting single-particle Fourier-like p.d.f. (\ref{eq:Fourier_cn_sn}). 

\begin{figure}[h]
\centering
\includegraphics[width=0.5\textwidth]{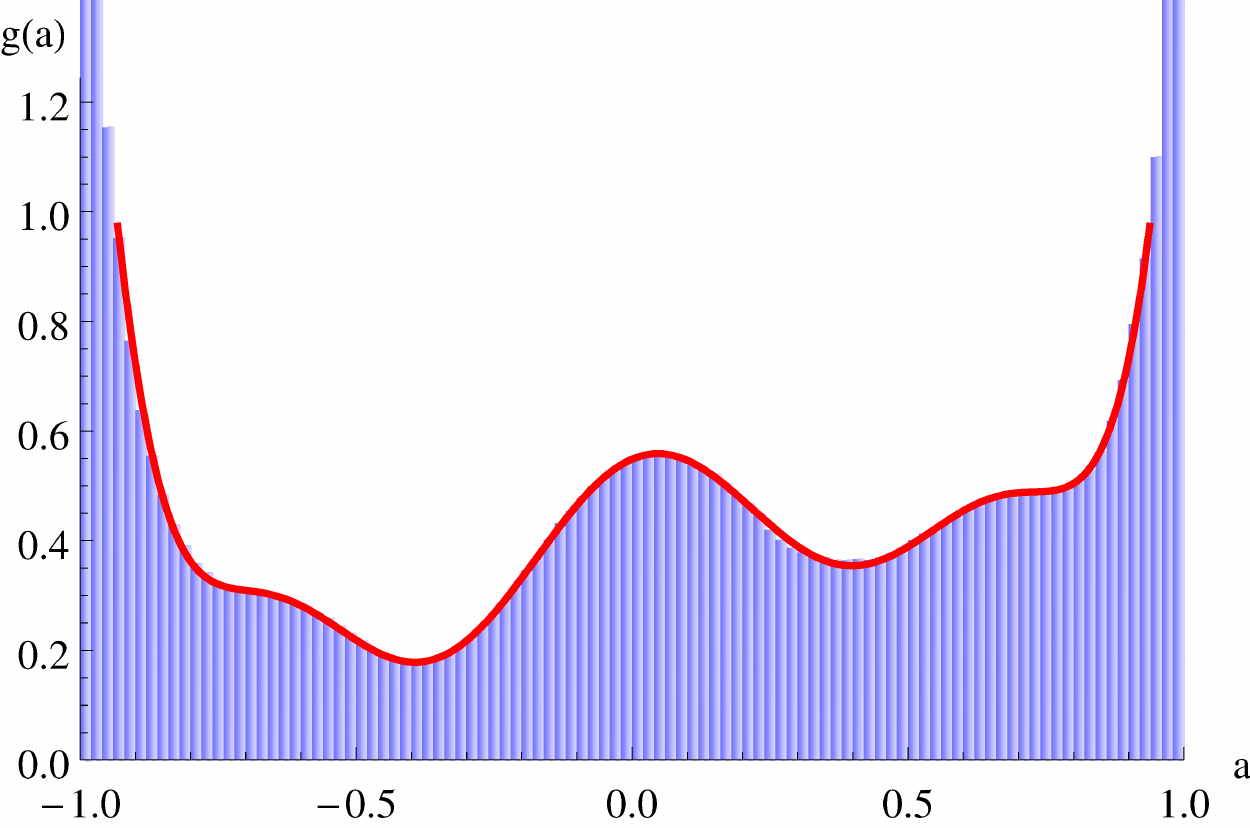}
\caption{(Color online) Distribution of the imaginary part of single-particle $Q$-vector (blue) and its theoretical p.d.f. (solid red line) obtained from Eq.~(\ref{eq:g(a)_fullFS_sinmx_secondVersion_main_part}) in a toy Monte Carlo example for the general case of multichromatic flow.}
\label{fig:generalCombinedCase_2d}
\end{figure}

Next we provide the analytic solutions for the p.d.f. of the imaginary part of single-particle $Q$-vector evaluated in harmonic $m$ for the most general case of multichromatic flow. This solution can be obtained straightforwardly by performing superpositions of fundamental solutions (\ref{eq:g(a)_Cos_Sin_main_part}) and (\ref{eq:g(a)_Sin_Sin_main_part}). Again, for completeness sake we summarize the starting setup which in this case is defined as:
\begin{equation}
x\in [0,2\pi),\qquad \displaystyle f(x)=\frac{1}{2\pi}\big[1+2\sum_{n=1}^{\infty}(c_n\cos nx+s_n\sin nx)\big],\qquad a(x)=\sin mx\,.
\end{equation}
For the p.d.f. $g(a)$ we have obtained the following analytic solution: 
\begin{equation}
g(a)=\frac{1}{\pi\sqrt{1-a^2}}\bigg(1+\displaystyle\sum_{\begin{subarray}{c}n\\(n/m\ {\rm even})\end{subarray}}^{\infty}\,2c_n\,i^{\frac{n}{m}}\,  T_{\frac{n}{m}}(a)+\displaystyle\sum_{\begin{subarray}{c}n\\(n/m\ {\rm odd})\end{subarray}}^{\infty}\,2s_n\,i^{\frac{n}{m}-1}\,  T_{\frac{n}{m}}(a)\bigg)\,,
\label{eq:g(a)_fullFS_sinmx_firstVersion_main_part}
\end{equation}
or written equivalently:
\begin{equation}
g(a)=\frac{1}{\pi\sqrt{1-a^2}}\bigg(1+\displaystyle\sum_{l=1}^{\infty}\,2(-1)^l(c_{2l\cdot m}T_{2l}(a) - s_{(2l-1)\cdot m}T_{2l-1}(a) )\bigg)\,.
\label{eq:g(a)_fullFS_sinmx_secondVersion_main_part}
\end{equation}
Given the result above, we see that the distribution of the imaginary part of a single-particle $Q$-vector is not sensitive to the presence of odd cosine terms in the Fourier series in Eq.~(\ref{eq:Fourier_cn_sn}). The analytic solution in Eq.~(\ref{eq:g(a)_fullFS_sinmx_secondVersion_main_part}) is illustrated with a toy Monte Carlo example shown on Fig.~\ref{fig:generalCombinedCase_2d}. In this example we have again randomly selected the values for ten harmonics $c_1,c_2,\ldots,c_{10}$ and for ten harmonics $s_1,s_2,\ldots,s_{10}$ from the interval $[0,0.1)$. We have used those 20 harmonics to parametrize the Fourier-like p.d.f. (\ref{eq:Fourier_cn_sn}). We remark again that the special care needs to be taken that the resulting Fourier-like p.d.f. is physical, i.e. positive definite over the whole interval $[0,2\pi)$, due to its probabilistic interpretation. From the sampled azimuthal angles we have calculated the imaginary part of the single-particle $Q$-vector, and created its distribution. For harmonic $m$ we have selected $m=1$, so that $a(x) = \sin x$ in this example. The resulting distribution of the imaginary part of the single-particle $Q$-vector is shown in blue on Fig.~\ref{fig:generalCombinedCase_2d}, while the theoretical result for $g(a)$ calculated from Eq.~(\ref{eq:g(a)_fullFS_sinmx_secondVersion_main_part}) is shown with the solid red line.

The results presented so far in this paper comprise the exact solutions for the p.d.f.'s of the real and imaginary parts of single-particle $Q$-vector for the most general case of the parametrization of initial Fourier-like p.d.f. (\ref{eq:Fourier_cn_sn}) with all possible harmonics $c_n$ and $s_n$. The next step is to obtain the p.d.f.'s of the real and imaginary parts of $M$-particle $Q$-vectors, from which finally the p.d.f's of multiparticle azimuthal correlations can be obtained. We make a progress in this direction in the next section.


\subsection{Characteristic functions}
\label{ss:Characteristic_functions}

For a random observable $a$ with the sample space $A$ and the p.d.f. $g(a)$, the characteristic function, denoted by $\phi_a(k)$, is defined as the inverse Fourier transform of its p.d.f. $g(a)$. In a lot of cases of practical interest working directly with characteristic functions provides additional insights both about the analytical and asymptotic properties of the problem in question. In the present study characteristic functions turned out to be very convenient because of the following property: The characteristic function for a sum of independent random observables is given by the product of the individual characteristics functions~\cite{Cowan:1998ji}. This is precisely the situation we have, because if the only source of correlations among produced particles is the collective anisotropic flow, then all particles are emitted independently to each other and are correlated only to some common symmetry planes. On the other hand, trivially from definitions (\ref{eq:Qvector}) and (\ref{eq:QvectorSingleParticleOrUnit}) we see that $M$-particle $Q$-vector is just a sum of $M$ single-particle $Q$-vectors. This means that it suffices to obtain characteristic functions of the single-particle $Q$-vectors, and then the characteristic function of $M$-particle $Q$-vector is trivially given by their product. In what follows we utilize the following definition for the characteristic function $\phi_a(k)$ for a random observable $a$ with a known p.d.f. $g(a)$~\cite{Cowan:1998ji}:
\begin{equation}
\phi_a(k) = \int_A\,e^{ika}g(a)\,da\,,
\label{eq:characteristicFunction_phi_a(k)}
\end{equation}
where the integration is performed over the whole sample space $A$ of $a$. In the present work we have assumed (see the discussion in Introduction) that particles are emitted from identical single-particle Fourier-like p.d.f.'s (\ref{eq:Fourier_cn_sn}), which yields to the following further simplification: The characteristic functions of real and imaginary parts of $M$-particle $Q$-vector is obtained from the characteristic functions of real and imaginary parts of the single-particle $Q$-vector raised to power $M$, respectively. 

\subsubsection{Characteristic functions for single-particle $Q$-vectors}
\label{ss:Characteristic-functions_for_single-particle_$Q$-vectors}

Characteristic functions both for the real and imaginary parts of single-particle $Q$-vectors defined in Eq.~(\ref{eq:QvectorSingleParticleOrUnit}) can be obtained analytically; we start to present our results with the simplest case of random walk. For some technical details which clarify the derivations of results to be presented in this section we refer to Appendix~\ref{app:Identities}. Taking into account the result for the p.d.f. $g(a)$ for random walk case in (\ref{eq:RW_main_part}) and the definition of characteristic function provided in (\ref{eq:characteristicFunction_phi_a(k)}), it follows immediately that the characteristic function of both the real and imaginary parts of single-particle $Q$-vector for the case of random walk is given by:
\begin{eqnarray}
\phi_a(k) &=& \int_{-1}^1\,\frac{e^{ika}}{\pi\sqrt{1-a^2}}\,da\nonumber\\
&=& J_0(k)\,,
\label{eq:characteristicFunction_RW_Case1_phi_a(k)}
\end{eqnarray}
where $J_0$ is Bessel function of the first kind and order zero, and either $a = \cos mx$ or $a = \sin mx$. 

Analytic results can be obtained also for the most general case of multichromatic flow. After some algebra, starting from the result for p.d.f. $g(a)$ in Eq.~(\ref{eq:g(a)_fullFS_cosmx_secondVersion_main_part}), utilizing identities from Appendix~\ref{app:Identities}, we have obtained that the characteristic function for the real part of single-particle $Q$-vector evaluated in harmonic $m$ is:
\begin{equation}
\phi_{{\rm Re}\,u_m}(k) = J_0(k) + 2\sum_{p=1}^\infty\,(-1)^p\left[c_{2p\cdot m}J_{2p}(k)
-  i c_{(2p-1)\cdot m} J_{2p-1}(k)\right]\,.
\label{eq:CharacteristicFunction_fullFS_cosmx}
\end{equation}
On the other hand, by using an analogous procedure we have obtained the following result for the characteristic function of the imaginary part of single-particle $Q$-vector evaluated in harmonic $m$:
\begin{equation}
\phi_{{\rm Im}\,u_m}(k) = J_0(k) + 2\sum_{p=1}^\infty\,\left[c_{2p\cdot m}J_{2p}(k)
+  i s_{(2p-1)\cdot m} J_{2p-1}(k)\right]\,.
\label{eq:CharacteristicFunction_fullFS_sinmx}
\end{equation}
Having obtained the above analytic results for the characteristic functions of single-particle $Q$-vectors, we trivially in the next section derive from them the results for the characteristic functions of $M$-particle $Q$-vectors, which then in turn can be used to obtain the analytic expressions for the higher order moments of the real and imaginary parts of $M$-particle $Q$-vectors. 

\subsubsection{Characteristic functions for $M$-particle $Q$-vectors}
\label{ss:Characteristic-functions_for_$M$-particle_$Q$-vectors}

As already indicated in the Introduction, under the assumption of factorization of a joint multivariate p.d.f.~(\ref{eq:factorization}) and under the assumption that the functional forms of each single-particle p.d.f. are the same and given by~(\ref{eq:Fourier_cn_sn}), the characteristic function for $M$-particle $Q$-vector can be obtained simply by raising the characteristic function of single-particle $Q$-vector to power $M$. Straightforwardly, for the case of random walk we have obtained: 
\begin{equation}
\phi_a(k) = J_0(k)^M\,.
\label{eq:CharacteristicFunction_RW_M-particle}
\end{equation}
where $a = {\rm Re}\,Q_m$ or $a = {\rm Im}\,Q_m$. 

On the other hand, for the case of multichromatic flow we obtain for the real part of $M$-particle $Q$-vector evaluated for harmonic $m$ the following analytic expression:
\begin{equation}
\phi_{{\rm Re}\,Q_m}(k) = \left[J_0(k) + 2\sum_{p=1}^\infty\,(-1)^p\left[c_{2p\cdot m}J_{2p}(k)
-  i c_{(2p-1)\cdot m} J_{2p-1}(k)\right]\right]^M\,,
\label{eq:CharacteristicFunction_Re_M-particle}
\end{equation}
while for the imaginary part of $M$-particle $Q$-vector evaluated for harmonic $m$ we have:
\begin{equation}
\phi_{{\rm Im}\,Q_m}(k) = \left[J_0(k) + 2\sum_{p=1}^\infty\,\left[c_{2p\cdot m}J_{2p}(k)
+  i s_{(2p-1)\cdot m} J_{2p-1}(k)\right]\right]^M\,.
\label{eq:CharacteristicFunction_Im_M-particle}
\end{equation}
Harmonics $c_n$ and $s_n$ in the above expression originate from the parametrization of initial single-particle Fourier-like p.d.f. in (\ref{eq:Fourier_cn_sn}). We see that the analytic expressions for characteristic functions of $M$-particle $Q$-vector are given solely in terms of Bessel functions of the first kind. From the knowledge of characteristic functions we can now obtain the p.d.f. $g(a)$ of both real and imaginary parts of $M$-particle $Q$-vectors by evaluating the following integral~\cite{Cowan:1998ji}:
\begin{equation}
g(a) = \frac{1}{2\pi}\int\phi_{a}(k)e^{-ika}dk\,,
\end{equation}
where either $a = {\rm Re}\,Q_m$ or $a = {\rm Im}\,Q_m$, and $\phi_{a}(k)$ is given by (\ref{eq:CharacteristicFunction_Re_M-particle}) and (\ref{eq:CharacteristicFunction_Im_M-particle}), respectively. From the results of p.d.f.'s of $M$-particle $Q$-vectors obtained in this way and the knowledge of analytic expressions which relate multiparticle azimuthal correlations with $M$-particle $Q$-vectors evaluated (in general) in different harmonics $m$, one can finally reach the final goal, i.e. the p.d.f.'s of multiparticle azimuthal correlations. We will present this in our parallel work~\cite{WorkInProgress}. 

In order to obtain, however, only the moments of the real and imaginary parts of $M$-particle $Q$-vectors the explicit knowledge of their p.d.f.'s is not needed~\cite{Cowan:1998ji}, and one can use only the characteristic functions (\ref{eq:CharacteristicFunction_Re_M-particle}) and (\ref{eq:CharacteristicFunction_Im_M-particle}) to determine analytically, in principle, all higher order moments. We carry out explicit calculations for few higher order moments of interest in this way in the next section and we use these results in the discussion on the sensitivity of multiparticle azimuthal correlations. 


\section{Higher order moments}
\label{s:Higher_order_moments}

If the characteristic function $\varphi_{a}(k)$ of a random observable $a$ is known, we can calculate the $n$th algebraic moment of $a$, to be denoted by $\mu'_{a,n}$, with the following expression~\cite{Cowan:1998ji}:
\begin{equation}
\mu'_{a,n} = i^{-n}\frac{d^n}{dk^n} \left.\phi_{a}(k)\right|_{k=0}\,.
\label{eq:mu'a_n_main_part}
\end{equation}
We will evaluate the above expression for the following cases of interest: $a={\rm Re}\,u_m$, $a={\rm Im}\,u_m$, $a={\rm Re}\,Q_m$ and $a={\rm Im}\,Q_m$, where $u_m$ was defined in Eq.~(\ref{eq:QvectorSingleParticleOrUnit}) while $Q_m$ was defined in Eq.~(\ref{eq:Qvector}). Results for the $n$th order algebraic moments of both the real and imaginary parts of single-particle $Q$-vector evaluated in harmonic $m$, to be denoted as $\mu'_{{{\rm Re}\,u_m},n}$ and $\mu'_{{{\rm Im}\,u_m},n}$ respectively, can be obtained analytically in a closed form (the technical details which led to the final results which we now present can be found in Appendix~\ref{app:Calculation_of_moments_from_characteristic_functions.}). In all results below $r$ denotes an arbitrary positive integer. 

For the $n$th algebraic moment of the real part of single-particle $Q$-vector evaluated in harmonic $m$ we have obtained for the most general case of multichromatic flow the following result:
\begin{equation}
\mu'_{{\rm Re}\,u_m,n}=
\left\{
 \begin{array}{ll}
\displaystyle
  \frac{(2r)!}{4^r(r!)^2}\bigg[1\!+\!2\sum_{p=1}^r c_{2p\cdot m}\frac{(r!)^2}{(r\!-\!p)!(r\!+\!p)!}\bigg]\,, &\qquad n=2r\,,\\
\displaystyle
    4^{1\!-\!r}\sum_{p=1}^r c_{(2p\!-\!1)\cdot m}\frac{(2r\!-\!1)!}{(r\!-\!p)!(r\!+\!p\!-\!1)!}\,, &\qquad n=2r\!-\!1\,, \\  
 \end{array} 
\right.
\label{eq:nth_moment_of_real_part_of_single-particle_Q-vector_evaluated_in_harmonic_m}
\end{equation}
while for the imaginary part we have:
\begin{equation}
\mu'_{{\rm Im}\,u_m,n}=
\left\{
 \begin{array}{ll}
\displaystyle
  \frac{(2r)!}{4^r(r!)^2}\bigg[1\!+\!2\sum_{p=1}^r c_{2p\cdot m}\frac{(-1)^p(r!)^2}{(r\!-\!p)!(r\!+\!p)!}\bigg]\,, & \qquad n=2r\,,\\
\displaystyle
    -4^{1\!-\!r}\sum_{p=1}^r s_{(2p\!-\!1)\cdot m}\frac{(-1)^p(2r\!-\!1)!}{(r\!-\!p)!(r\!+\!p\!-\!1)!}\,, &\qquad n=2r\!-\!1\,. \\  
 \end{array} 
\right.
\label{eq:nth_moment_of_imaginary_part_of_single-particle_Q-vector_evaluated_in_harmonic_m}
\end{equation}
Although for the general case of multichromatic flow there is an infinite number of terms in the initial single-particle Fourier-like p.d.f. in Eq.~(\ref{eq:Fourier_cn_sn}), and in the characteristic functions in Eqs.~(\ref{eq:CharacteristicFunction_fullFS_cosmx}) and (\ref{eq:CharacteristicFunction_fullFS_sinmx}), we see that there is a {\it finite} number of terms which can contribute to algebraic moments of single-particle $Q$-vectors in Eqs.~(\ref{eq:nth_moment_of_real_part_of_single-particle_Q-vector_evaluated_in_harmonic_m}) and (\ref{eq:nth_moment_of_imaginary_part_of_single-particle_Q-vector_evaluated_in_harmonic_m}) (see Appendix~\ref{app:Calculation_of_moments_from_characteristic_functions.} for a detailed justification). This observation implies that each individual component of multichromatic flow can be independently studied and experimentally constrained via the finite number of suitably chosen algebraic moments.  

The analytic results for the algebraic moments of the real and imaginary parts of $M$-particle $Q$-vectors  can be obtained straightforwardly by inserting the results for characteristic functions from Eqs.~(\ref{eq:CharacteristicFunction_Re_M-particle}) and (\ref{eq:CharacteristicFunction_Im_M-particle}) into Eq.~(\ref{eq:mu'a_n_main_part}). We have obtained the following analytic results for the first four algebraic moments of $M$-particle $Q$-vectors evaluated in harmonic $m$:
\begin{eqnarray}
\mu'_{{{\rm Re}\,Q_m},1} &=& Mc_m\,,\\ 
\mu'_{{{\rm Im}\,Q_m},1} &=& Ms_m\,,\\ 
\mu'_{{{\rm Re}\,Q_m},2} &=& \frac{M}{2}\bigg[1\!+\!2(M\!-\!1)c_m^2\!+\!c_{2m}\bigg]\label{eq:mu'req_2}\,,\\ 
\mu'_{{{\rm Im}\,Q_m},2} &=& 
\frac{M}{2}\bigg[1\!+\!2(M\!-\!1)s_m^2\!-\!c_{2m}\bigg]\label{eq:mu'imq_2}\,,\\ 
\mu'_{{{\rm Re}\,Q_m},3} &=&\frac{M}{4}\bigg[4(M\!-\!2)(M\!-\!1) c_m^3\!+\!3(2M\!-\!1)c_m\!+\!6 (M\!-\!1)c_{m}c_{2m}\!+\!c_{3m}\bigg]\,,\\ 
\mu'_{{{\rm Im}\,Q_m},3} &=& \frac{M}{4}\bigg[4(M\!-\!2)(M\!-\!1) s_m^3\!+\!3(2M\!-\!1)s_m\!-\!6 (M\!-\!1)s_{m}c_{2m}\!-\!s_{3m}\bigg]\,,\\
\mu'_{{{\rm Re}\,Q_m},4} &=& \frac{M}{8} \bigg[3(2M\!-\!1)\!+\!8(M\!-\!3)(M\!-\!2)(M\!-\!1)c_m^4\!+\!24 (M\!-\!1)^2 c_m^2\!+\!6(M\!-\!1)c_{2m}^2\!\nonumber\\
&&{}+4(3M\!-\!2)c_{2m}\!+\!c_{4m}\!+\!24(M\!-\!1)(M\!-\!2)c_m^2c_{2m}\!+\!8(M\!-\!1)c_m c_{3m}\bigg]\,,\\ 
\mu'_{{{\rm Im}\,Q_m},4} &=&\frac{M}{8}\bigg[3(2M\!-\!1)\!+\!8(M\!-\!3)(M\!-\!2)(M\!-\!1)s_m^4\!+\!24 (M\!-\!1)^2 s_m^2\!+\!6(M\!-\!1)c_{2m}^2\!\nonumber\\
&&{}-4(3M\!-\!2)c_{2m}\!+\!c_{4m}\!-\!24(M\!-\!2)(M\!-\!1)s_m^2c_{2m}\!-\!8(M\!-\!1)s_m s_{3m}\bigg]\,. 
\label{eq:algebraic_moments}
\end{eqnarray}
All above moments are well defined observables and can be obtained independently and directly only  from the azimuthal angles of produced particles in each heavy-ion collision. For the case when the initial single-particle p.d.f. was parametrized with the Fourier series as in Eq.~(\ref{eq:Fourier_cn_sn}), the strict functional forms of the above moments, with harmonics $c_n$ and $s_n$ considered as variables, and their mutual relations carry the imprint of full factorization (\ref{eq:factorization}) of joint multivariate p.d.f. We stress again that such a full factorization occurs if the correlations among produced particles are dominated by correlations originating from collective anisotropic flow. This leads us to our key point: The strict functional forms of above moments can be used as a hypothesis to test whether the observed event-by-event anisotropies in heavy-ion collisions are originating from collective anisotropic flow. On the other hand, the functional form of a single-particle p.d.f. in Eq.~(\ref{eq:Fourier_cn_sn}) by itself quantifies any type of anisotropy in the particle production, whether or not it originates from the collective anisotropic flow. 

Due to random fluctuations of impact parameter vector, which experimentally cannot be controlled, the above algebraic moments cannot be combined straightforwardly for an ensemble of heavy-ion collisions in order to suppress the statistical spread in their measurements---therefore their direct usage is limited only to theoretical studies, when the orientation of impact parameter vector in each heavy-ion collision can be fully controlled. This issue can be fully overcomed experimentally by considering the algebraic moments of isotropic azimuthal observables, i.e. observables which are invariant under the transformation $\varphi\mapsto\varphi+\alpha$, where $\varphi$ labels azimuthal angles of produced particles and $\alpha$ is arbitrary. For instance, such observables are isotropic two- and multiparticle azimuthal correlations, and we focus next on them. 

In some special cases the algebraic moments of isotropic azimuthal correlations can be obtained directly from the algebraic moments of $M$-particle $Q$-vectors discussed above. As an example, from the analytic relation which expresses isotropic two-particle azimuthal correlation in terms of $M$-particle  $Q$-vectors~\cite{Bilandzic:2010jr},
\begin{eqnarray}
\left<2\right>_{m,-m} &=& \frac{1}{M(M\!-\!1)}\,\sum_{\begin{subarray}{c}i,j=1\\i\neq j\end{subarray}}^{M}\cos m(\varphi_i-\varphi_j)\nonumber\\
&=&\frac{|Q_m|^2\!-\!M}{M(M\!-\!1)}\,,
\label{eq:2pBasicExample_first_part}
\end{eqnarray}
it follows that the first algebraic moment (the mean) of isotropic two-particle azimuthal correlation is given as:
\begin{eqnarray}
\mu_{\left<2\right>_{m,-m}} &=& \frac{\mu'_{|Q_m|^2,1}-M}{M(M\!-\!1)}\nonumber\\
&=&\frac{\mu'_{{{\rm Re}\,Q_m},2}\!+\!\mu'_{{{\rm Im}\,Q_m},2}\!-\!M}{M(M\!-\!1)}\nonumber\\
&=&c_m^2+s_m^2\nonumber\\
&=&v_m^2\,,
\label{eq:2pBasicExample_second_part}
\end{eqnarray}
where into the second line above we have inserted the results (\ref{eq:mu'req_2}) and (\ref{eq:mu'imq_2}), while in order to get the last line we have just switched to standard parametrization of Fourier series in terms of amplitudes $v_n$ and symmetry planes $\Psi_n$. In general, however, additional algebraic manipulations are required in order to obtain moments of isotropic multiparticle azimuthal correlations from the  
moments of real and imaginary parts of $M$-particle $Q$-vectors. In a self-contained Appendix~\ref{app:Moments} we have established one such procedure, which can be used in the derivation of moments of cross-terms, i.e. the terms which depend on products of the real and imaginary parts of $M$-particle $Q$-vectors. For simplicity, in the next section we use only the simplified results applicable in the {\it heavy-ion regime} for the higher order moments of isotropic two- and four-particle azimuthal correlations in the discussion on sensitivity of anisotropic flow measurements obtained with correlation techniques. 


\subsection{Sensitivity}
\label{ss:Sensitivity}

We define the heavy-ion regime to be the regime of large multiplicities, $M\!>\!100$, and in which one harmonic is dominant and has magnitude of the order of $v \sim 0.05$, while all other harmonics are much smaller in magnitude. Under such assumptions, the expressions for the variance of isotropic two- and four-particle correlations evaluated for the dominant harmonic, to be denoted $\sigma_{\left<2\right>}^2$ and $\sigma_{\left<4\right>}^2$ respectively, simplify tremendously. We obtain for the heavy-ion regime:
\begin{eqnarray}
\sigma_{\left<2\right>}^2 &\simeq& \frac{1}{M^2}\left(1\!+\!2\chi^2\right)\,,\label{eq:sigma2HI}\\
\sigma_{\left<4\right>}^2 &\simeq& \frac{4}{M^4}\left(1\!+\!4\chi^2\!+\!5\chi^4\!+\!2\chi^6\right)\,.\label{eq:sigma4HI}
\end{eqnarray}
The approximate result~(\ref{eq:sigma2HI}) was obtained from the analytic result~(\ref{eq:variance_Appendix}), while the approximate result (\ref{eq:sigma4HI}) was obtained from the analytic result originally first calculated in~\cite{Peter_Bachelor_Project}. We have in the above expressions introduced $\chi^2\equiv Mv^2$ as a resolution parameter. These approximate expressions (\ref{eq:sigma2HI})
and (\ref{eq:sigma4HI}) were tested in a toy Monte Carlo study and it was found that they have accuracy better than 5\% for the dominant harmonic in the heavy-ion regime described above. 

The final result for isotropic two-particle azimuthal correlation, $\left<2\right>$, is reported as:
\begin{equation}
\mu_{\left<2\right>} \pm \frac{\sigma_{\left<2\right>}}{\sqrt{N}}\,,
\end{equation}
where $N$ is total number of events. If we want the statistical noise to be suppressed at the level or better than $n$ significant digits, we impose the following constraint:
\begin{equation}
\frac{\rm statistical\ noise}{\rm result} < 10^{-n}
\\,
\end{equation}
which translates into
\begin{equation}
\frac{\sigma_{\left<2\right>}}{\mu_{\left<2\right>}\sqrt{N}} < 10^{-n}\,.
\end{equation}
\begin{figure*}[t]
  \begin{center} 
\includegraphics[width=0.60\textwidth]{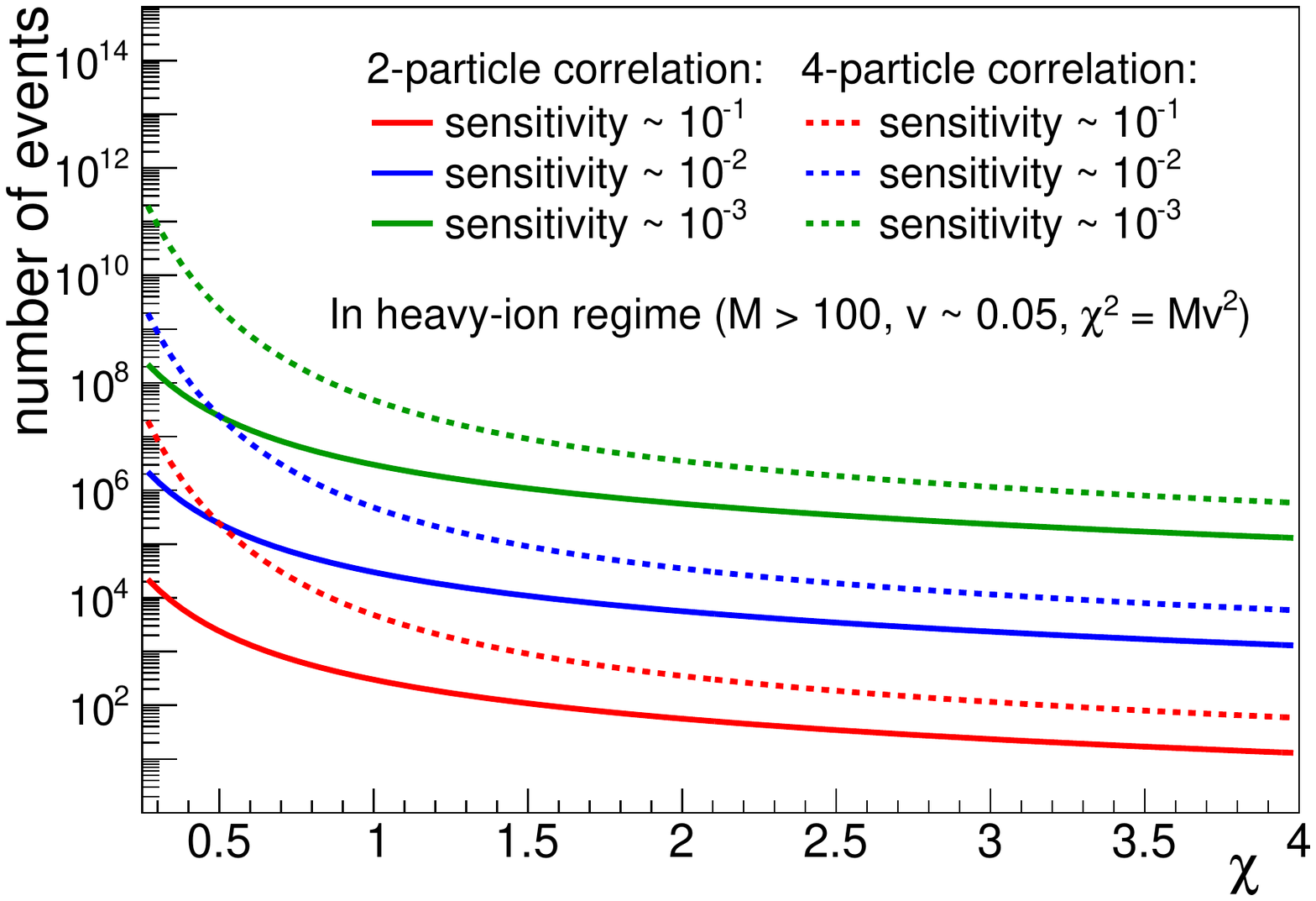}
    \caption{Sensitivity of isotropic two-particle (solid lines) and four-particle (dashed lines) azimuthal correlations in a heavy-ion regime as a function of resolution parameter $\chi$, which is defined as $\chi^2=Mv^2$, where $M$ is the multiplicity and $v$ is the flow harmonic. The heavy-ion regime is defined as the regime of large multiplicities, $M > 100$, and in which one harmonic is dominant and has magnitude of the order of $v \sim 0.05$, while all other harmonics are much smaller in magnitude. We see from the figure that for a typical values of resolution in heavy-ion collisions at Large Hadron Collider, namely $\chi\sim 1.5$, we need more than 1 million events to suppress statistical noise beyond three significant digits, both in the measurements of two- and four-particle azimuthal correlations.} 
    \label{fig:sensitivity_0b}
  \end{center}
\end{figure*}

\noindent Taking into account the results obtained for the mean and variance of $\left<2\right>$, we have after some algebra for the heavy-ion regime:
\begin{equation}
N > 100^n \frac{1+2\chi^2}{\chi^4}\,,\qquad\chi^2=Mv^2\,.
\label{eq:Beyond_n_HI_2p}
\end{equation}
By following a completely analogous procedure, we have obtained the following estimate for the required number of events for the case of isotropic four-particle correlation:
\begin{equation}
N > 100^n \frac{4(1+4\chi^2+5\chi^4+2\chi^6)}{\chi^8}\,,\qquad\chi^2=Mv^2\,.
\label{eq:Beyond_n_HI_4p}
\end{equation}
The relations (\ref{eq:Beyond_n_HI_2p}) and (\ref{eq:Beyond_n_HI_4p}) indicate, under the assumptions discussed above, how many events at least have to be collected in order for statistical noise to be suppressed beyond $n$ significant digits in the measurement of isotropic two- and four-particle azimuthal correlations, respectively (for a detailed discussion on sensitivity of correlation techniques when the formalism of generating functions is utilized, we refer to Appendix~D of~\cite{Borghini:2001vi}). We have illustrated the above two estimates for the number of events in Fig.~\ref{fig:sensitivity_0b} as a function of resolution parameter $\chi$. It can be easily concluded from the above results that for all values of resolution $\chi$ bigger than 1.5 we need about 4 times more events in the measurements of four-particle correlations in order to make them equally sensitive to the measurements of two-particle correlations. Flattening at large $\chi$ indicates that neither two- nor four-particle azimuthal correlations are suitable in their present form for the direct event-by-event flow analysis.  

As our last result, we report the result for the skewness (as defined in Eq.~(\ref{eq:skewness})) of isotropic two-particle azimuthal correlation. The exact result can be obtained directly from the moments of $Q$-vector amplitudes in Appendix~\ref{app:Exact_results_for_the_moments}, here we present only the approximate result applicable in the heavy-ion regime:
\begin{equation}
\gamma_{1,\left<2\right>}\simeq\frac{2(1\!+\!3\chi^2)}{(1\!+\!2\chi^2)^{3/2}}\,,\qquad\chi^2=Mv^2\,.
\label{eq:skewness_approximate_result}
\end{equation}
The accuracy of the above approximate result is better than 5\% in the heavy-ion regime. The non-zero value for the skewness indicates an asymmetric non-Gaussian nature of the underlying p.d.f., while its positive value indicates that the tail of underlying p.d.f. is longer on the right side than on the left side.


\section{Summary}
\label{s:Summary}

We have derived the analytic expressions in the most general case of multichromatic flow for the p.d.f.'s of the real and imaginary parts of single-particle $Q$-vectors and demonstrated that they can be expressed solely in terms of Chebyshev polynomials of the first kind. From these results the analytic expressions for the characteristic functions of the real and imaginary parts of $M$-particle $Q$-vectors were derived and expressed in terms of Bessel functions of the first kind. From these analytic expressions for the characteristic functions all moments of the real and imaginary parts of $M$-particle $Q$-vector can be obtained, and we have provided explicit results for the first four algebraic moments. We have used results for the variance of isotropic two and four-particle azimuthal correlations in the discussion of sensitivity of anisotropic flow measurements with correlation techniques. The present paper paves the road towards our ultimate goal: The derivation of analytic expressions for the probability density functions of multiparticle azimuthal correlations.


\acknowledgments{
We thank Jens J\o rgen Gaardh\o je for encouraging comments and for providing valuable feedback on the paper.  We thank Kristjan Gulbrandsen for fruitful discussions. We thank the other members of HEHI group at NBI for showing growing interest in this project. We thank Jiangyong
Jia for a useful feedback on the event-by-event flow measurements via unfolding methods.  We thank the Danish Council for Independent Research, Natural Sciences (FNV) and the Danish National Research Foundation (Danmarks Grundforskningsfond) for support.}

\appendix

\section{Distributions of functions of random observables}
\label{app:Distributions_of_functions_of_random_observables}


In the first part of this appendix we present the general procedure which will be used in this project in order to determine the p.d.f. of a function of random observable with known p.d.f. The presentation is heavily motivated by the basic material provided in the classical textbook~\cite{Cowan:1998ji}, and is further generalized here in order to cover all cases of our interest which were encountered in this study. In the second part of the appendix we demonstrate how this general procedure can be used by working out in detail the toy Monte Carlo example. Finally, in the third part we prove a few general claims which were helpful in the derivations of our main results.   

\subsection{General procedure}
\label{app:General_procedure}

Our starting point is a continuous random observable $x$ whose sample space we denote by $X$, and whose behavior is governed by the p.d.f. $f(x)$. Any function $a(x)$ is then also a random observable, and we are interested in finding its p.d.f. $g(a)$ once  $X$, $f(x)$ and $a(x)$ are all specified. The sample space of $a$ we denote by $A$. From the conservation of probability we have~\cite{Cowan:1998ji}:
\begin{equation}
g(a)da = f(x)dx\,.
\label{eq:ConservationOfProbability}
\end{equation}
This result, however, can be applied directly in the above form only in some rather limited and simple cases---we now elaborate on all steps which are needed in the more general cases which will be encountered in this study. 

In the first step we fragment the sample space $X$ of random observable $x$ into $N$ disjoint subsets $X_1, X_2,\ldots,X_N$, which are of a different length in general. The fragmentation is being performed until for each resulting subset $X_i$ all of the following requirements are satisfied:
\begin{enumerate}
\item $a(x)$ has a unique functional form over $X_i$; 
\item $a(x)$ has a unique inverse, denoted by $x(a)$, over $X_i$;
\item $f(x)$ has a unique functional form over $X_i$.
\end{enumerate}
The first and third requirements above originate from the fact that in general both $a(x)$ and $f(x)$ can be piecewise-defined, while the second requirement indicates that the special treatment is needed for instance for periodic functions $a(x)$. We remark that for a piecewise-defined p.d.f. $f(x)$ one has to specify separately both the probability that observable will be sampled in a certain subinterval and the functional form of p.d.f. which applies in that subinterval. We utilize the following convention for the definition of piecewise-defined p.d.f. $f(x)$: 
\begin{equation}
f(x)\equiv
\left\{
 \begin{array}{lc}
  p_{X_{1}}f_{X_{1}}(x)\,, & x \in X_1\,, \\ 
  p_{X_{2}}f_{X_{2}}(x)\,, & x \in X_2\,, \\ 
  \ldots&\\
  p_{X_{N}}f_{X_{N}}(x)\,, & x \in X_N\,,  
 \end{array} 
\right.
\label{eq:f(x)_piecewise_defined}
\end{equation}
where $X=X_1\cup X_2\cup\cdots\cup X_N$. The p.d.f. $f(x)$ is normalized to unity over the whole sample space $X$, i.e.
\begin{equation}
\int_X f(x)\,dx = 1\,,
\label{eq:f(x)_normalization_to_unity}
\end{equation}
and in the definition (\ref{eq:f(x)_piecewise_defined}) we impose the constraints
\begin{equation}
\sum_{i=1}^N\,p_{X_{i}} = 1\,,
\label{eq:sum_of_probabilties}
\end{equation}
and
\begin{equation}
\int_{X_{i}}f_{X_{i}}(x)\,dx = 1\,,\qquad \forall i\,.
\label{eq:subinterval_normalization}
\end{equation}
Taking into account all above specifications, $p_{X_i}$ is the probability that observable $x$ will be sampled in the subinterval $X_i$, while $f_{X_i}(x)$ is a normalized p.d.f. over subinterval $X_i$ which completely determines the behaviour of observable $x$ within the subinterval $X_i$.
 
After the fragmentation of $X$ has been finalized, in the second step we map the boundaries of each resulting subinterval $X_i$ with the functional form of $a(x)$ applicable over that subinterval. This mapping results in the set of values $a_1,a_2,a_3,\ldots$ in the sample space $A$ of a random observable $a$. It is important to realize that each boundary at which $a(x)$ has a discontinuity must be mapped with functional forms of $a(x)$ which are applicable on both sides of discontinuity. The resulting set $a_1,a_2,a_3,\ldots$ has to be ordered, and such an ordered set will define the boundaries of disjoint subsets $A_1,A_2,A_3,\ldots$ in the sample space $A$. Finally, for each subset $A_j$ we collect all disjoint subsets of $X$ which were mapped with $a(x)$ into $A_j$, and we label them as $X_{j1}, X_{j2}, X_{j3},\ldots$. Then in each disjoint subset $A_j$ the probability to observe random observable $a$ in the interval $[a,a\!+\!da)$ is given by the following expression:  
\begin{equation}
g_j(a)da = \sum_i p_{ji}\, g_{ji}(a) da\,.
\label{eq:g(a)_combined}
\end{equation}
Index $j$ in the above equation labels the certain subset of $A$, while index $i$ runs over all subsets of $X$ which were mapped with $a(x)$ into the subset of $A$ labeled by $j$. Definition and interpretation of probability factors $p_{ji}$ are provided in Eq.~(\ref{eq:p_ji}) further below. With $g_{ji}(a)$ in  Eq.~(\ref{eq:g(a)_combined}) we have denoted the p.d.f. of $a$ when $x$ is restricted to belong only to the subset $X_{ji}$. From~\cite{Cowan:1998ji} we have:
\begin{equation}
g_{ji}(a) = f_{ji}(x(a))\left|\frac{dx}{da}\right|\,,
\label{eq:g_i(a)}
\end{equation}
where $f_{ji}(x)$ is a normalized p.d.f. of random observable $x$ over subinterval $X_{ji}$ which is obtained from the functional form of the starting $f(x)$ (see generic definition in Eq.~(\ref{eq:f(x)_piecewise_defined})) applicable in the subinterval $X_{ji}$. On the other hand, $x(a)$ is the inverse of $a(x)$. Since in general $a(x)$ is piecewise-defined, one has to take special care here to take into account the functional form of $a(x)$ which applies over subinterval $X_{ji}$, and in the case of periodic functions $a(x)$ also the correct branch of the inverse function $x(a)$ in the subinterval $X_{ji}$. Finally, given the generic definition of $f(x)$ in Eq.~(\ref{eq:f(x)_piecewise_defined}), the probability factors $p_{ji}$ in Eq.~(\ref{eq:g(a)_combined}) are determined as: 
\begin{equation}
p_{ji} \equiv \frac{\int_{X_{ji}}f(x)\,dx}{\sum_i \int_{X_{ji}}f(x)\,dx}\,.
\label{eq:p_ji}
\end{equation}
With the above definition, $p_{ji}$ is the probability that $x$ will be sampled in the subset $X_{ji}$ if the total sample space of $x$ is restricted only to $X_{j1}\cup X_{j2}\cup\cdots$. Looking from another angle, if the total sample space of $a$ is restricted only to $A_j$, then $p_{ji}$ is the probability that $a\in A_j$ was obtained with $a(x)$ where $x\in X_{ji}$. The definition (\ref{eq:p_ji}) ensures that the resulting p.d.f. of $a$ in subset $A_j$, namely $g_j(a)$ in Eq.~(\ref{eq:g(a)_combined}), is automatically normalized to unity over the subset $A_j$. Our final result for the p.d.f. $g(a)$ for the whole sample space $A=A_1\cup A_2\cup \cdots$ of a function $a(x)$ will in general be reported as:
\begin{equation}
g(a)=
\left\{
 \begin{array}{lc}
  p_1\,g_1(a)\,, & a \in A_1\,, \\ 
  p_2\,g_2(a)\,, & a \in A_2\,, \\ 
  \ldots
 \end{array} 
\right.
\label{eq:g(a)_definition_final_result}
\end{equation}
where each $g_j(a)$ above is calculated separately with Eq.~(\ref{eq:g(a)_combined}) and is normalized to unity by construction over the subset $A_j$. On the other hand, $p_j$ denotes the probability that the sampled value of $x$ from the starting p.d.f. $f(x)$ is mapped further with $a(x)$ in the subset $A_j$; it is given by:
\begin{equation}
p_j\equiv\sum_i \int_{X_{ji}}f(x)\,dx\,, 
\label{eq:p_j}
\end{equation}
where $f(x)$ is the starting p.d.f. of $x$ defined in Eq.~(\ref{eq:f(x)_piecewise_defined}), while the summation and integration are performed over all subsets of $X$ which were mapped with $a(x)$ into the subset $A_j$ of the sample space $A$. By construction, the final p.d.f. $g(a)$ written as in (\ref{eq:g(a)_definition_final_result}) is automatically normalized to unity over the whole sample space $A$.

For clarity sake, we now go through all above steps in a detailed toy Monte Carlo example.  

\subsection{Toy Monte Carlo example}
\label{app:Toy_Monte_Carlo_example}

\begin{figure}[h]
\centering
\includegraphics[width=0.5\textwidth]{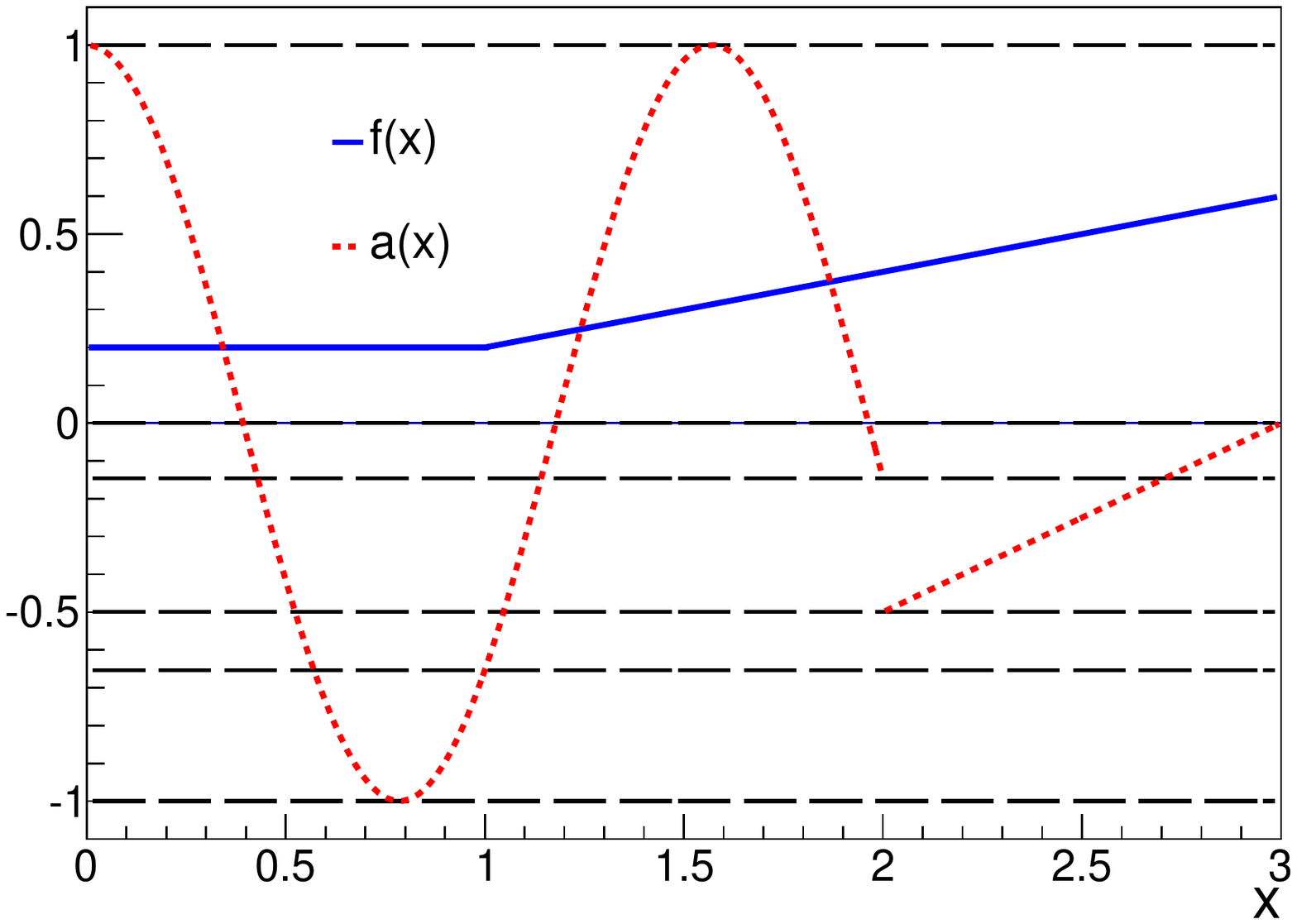} 
\caption{(Color online) Piecewise-defined p.d.f. $f(x)$ from definition (\ref{eq:pdf_f(x)_MC}) and a piecewise-defined function $a(x)$ from definition (\ref{eq:a(x)_MC}) of a random observable $x$, whose sample space is the interval $[0,3)$. For the explanation of horizontal dashed lines see the main text.}
\label{fig:f_and_a_WithIntervals_0c}
\end{figure}

\begin{figure}[h]
\centering
\includegraphics[width=0.5\textwidth]{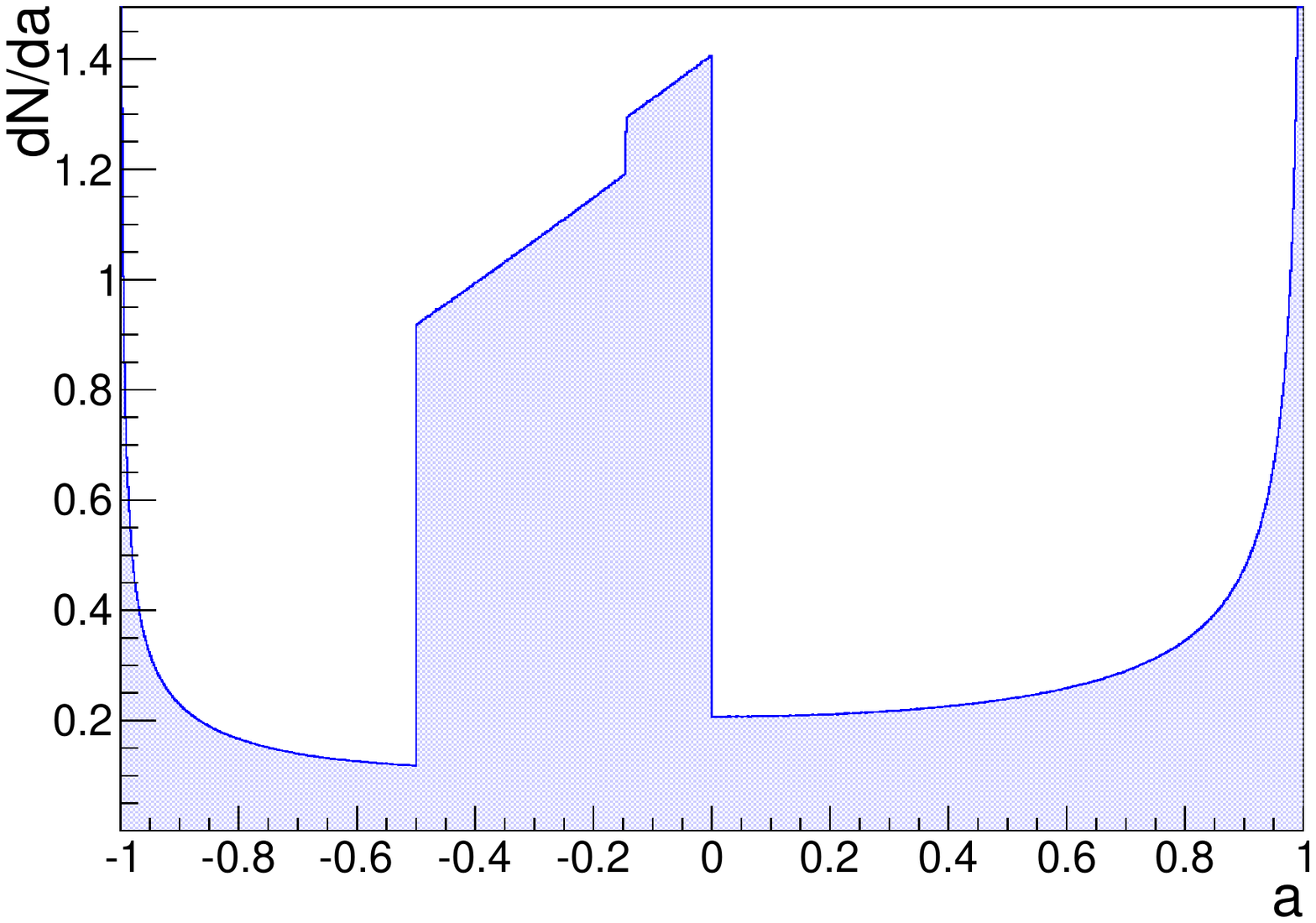}
\caption{(Color online) The resulting distribution of a random observable $a$ in a toy Monte Carlo example.}
\label{fig:distribution_of_a_0c}
\end{figure}

In this section we set up the toy Monte Carlo example in order to illustrate all the steps which are required in the derivation of a p.d.f. of a function of random observable with known p.d.f. This example is general enough to cover all distinct cases which we will encounter within the scope of this project. The starting p.d.f. $f(x)$ of a random observable $x$ with sample space $X\equiv[0,3)$ is a piecewise-defined function as follows:
\begin{equation}
f(x)\equiv
\left\{
 \begin{array}{lc}
  \frac{1}{5}\cdot 1\,, & x \in [0,1)\,, \\ 
  \frac{4}{5}\cdot \frac{1}{4}\,x\,, & x \in [1,3)\,.  
 \end{array} 
\right.
\label{eq:pdf_f(x)_MC}
\end{equation}
The above definition indicates that one first with probabilities $p_{X_1}=\frac{1}{5}$ and $p_{X_2}=\frac{4}{5}$ determines the interval $X_1=[0,1)$ or $X_2=[1,3)$, respectively, after which the sampling is performed within the chosen interval either with $f_{X_1}(x)=1$ or $f_{X_2}(x)=\frac{1}{4}\,x$. Both $f_{X_1}(x)$ and $f_{X_2}(x)$ are normalized to unity over intervals $X_1=[0,1)$ and $X_2=[1,3)$, respectively. 
This, together with the fact that $p_{X_1}+p_{X_2}=1$, ensures that $f(x)$ is normalized to unity over the whole sample space $X=[0,3)$ (see Eqs.~(\ref{eq:f(x)_normalization_to_unity}), (\ref{eq:sum_of_probabilties}) and (\ref{eq:subinterval_normalization})). On the other hand, the function $a(x)$ is defined as the following piecewise function:
\begin{equation}
a(x)\equiv
\left\{
 \begin{array}{lc}
  \cos 4x\,, & x \in [0,2)\,, \\ 
  \frac{1}{2}(x-3)\,, & x \in [2,3)\,.  
 \end{array} 
\right.
\label{eq:a(x)_MC}
\end{equation}
The above setup completely determines our toy Monte Carlo problem. We sample random observable $x$ from the p.d.f. $f(x)$ defined in Eq.~(\ref{eq:pdf_f(x)_MC}) and from the sampled value of $x$ we calculate random observable $a$ from the definition of  $a(x)$ in (\ref{eq:a(x)_MC}). On Fig.~\ref{fig:f_and_a_WithIntervals_0c} we have shown both $f(x)$ (solid blue line) and $a(x)$ (dotted red line) used in this toy Monte Carlo example, while the resulting distribution of $a$ is shown in Fig.~\ref{fig:distribution_of_a_0c}. Our aim is to derive analytically the p.d.f. $g(a)$ which describes the distribution of $a$ shown in Fig.~\ref{fig:distribution_of_a_0c}.

We first proceed with the fragmentation of the sample space $X$ of the starting random observable $x$. The first requirement (see Section~\ref{app:General_procedure} for the full list of requirements), namely that $a(x)$ has a unique functional form within each subset of $X$ under consideration, results in $X=X_1\cup X_2$, where $X_1 = [0,2)$ and $X_2 = [2,3)$. The next requirement that $a(x)$ has a unique inverse within each subset $X_i$ results in further fragmentation of subset $X_1$ from previous step, so that now $X=X_1\cup X_2\cup X_3\cup X_4$, where $X_1 = [0,\frac{\pi}{4})$, $X_2 = [\frac{\pi}{4},\frac{\pi}{2})$, $X_3 = [\frac{\pi}{2},2)$ and $X_4 = [2,3)$. Finally, the requirement that $f(x)$ must have a unique functional form over each subset $X_i$ results in the fragmentation of subset $X_2$ from the previous step, so that after this requirement we have reached the final fragmentation as $X=X_1\cup X_2\cup X_3\cup X_4\cup X_5$, where now $X_1 = [0,\frac{\pi}{4})$, $X_2 = [\frac{\pi}{4},1)$, $X_3 = [1,\frac{\pi}{2})$, $X_4 = [\frac{\pi}{2},2)$ and $X_5 = [2,3)$.

In the second step we map the boundaries of each subset $X_i$, $1 \leq i \leq 5$, in the final fragmentation with the functional form of $a(x)$ which applies over that subset. It follows:
\begin{eqnarray}
0&\mapsto& 1\,,\nonumber\\
\pi/4&\mapsto& -1\,,\nonumber\\
1&\mapsto& \cos 4 \simeq -0.654\,,\nonumber\\
\pi/2&\mapsto& 1\,,\nonumber\\
2&\mapsto& \cos 8 \simeq -0.146,\ -1/2\,,\nonumber\\
3&\mapsto& 0\,.
\label{eq:mapping}
\end{eqnarray}
We remark that the point $x=2$ has been mapped into two points in the sample space $A$, namely $a=-1/2$ and $a=\cos 8 \simeq -0.146$, due to the fact that $a(x)$ has a discontinuity at $x=2$ (i.e. we had to use functional forms of $a(x)$ which apply on both sides of discontinuity). Finally, we order the points in (\ref{eq:mapping}) to obtain sequence $-1,-0.654,-0.5,-0.146,0,1$, from which we read off the disjoint subsets $A_j$ in $A$:
\begin{eqnarray}
A_1 &=& [-1,-0.654)\,,\nonumber\\
A_2 &=& [-0.654,-0.5)\,,\nonumber\\
A_3 &=& [-0.5,-0.146)\,,\nonumber\\
A_4 &=& [-0.146,0)\,,\nonumber\\
A_5 &=& [0,1)\,.
\label{eq:Ai's}
\end{eqnarray}
The boundaries of above subsets of $A$ are indicated with horizontal dashed lines in Fig.~\ref{fig:f_and_a_WithIntervals_0c}. We seek solutions for the p.d.f. of $a(x)$ in each subset in (\ref{eq:Ai's}) separately. We provide detailed calculation below only for the subset $A_2$, while for all other subsets we provide only the final result for p.d.f. $g_i(a)$ in the subset $A_i$. 

There are two subintervals (see Fig.~\ref{fig:f_and_a_WithIntervals_0c}), namely $X_{21}=[\frac{\pi}{6},\frac{\pi}{2}\!-\!1)$ and $X_{22}=[1,\frac{\pi}{3})$ which are mapped with $a(x)$ defined in (\ref{eq:a(x)_MC}) into $A_2=[-0.654,-0.5)$. First we normalize the starting p.d.f. $f(x)$ defined in (\ref{eq:pdf_f(x)_MC}) to unity over interval $X_{21}=[\frac{\pi}{6},\frac{\pi}{2}\!-\!1)$ to obtain normalized p.d.f. $f_{X_{21}}(x)$ in that interval: 
\begin{equation}
f_{X_{21}}(x) = \frac{3}{\pi\!-\!3}\,,\qquad x \in [\frac{\pi}{6},\frac{\pi}{2}\!-\!1)\,. \\ 
\label{eq:f(x)_X21}
\end{equation}
Analogously, the normalized p.d.f. $f_{X_{22}}(x)$ in the interval $X_{22}=[1,\frac{\pi}{3})$ is: 
\begin{equation}
f_{X_{22}}(x)=\frac{18}{\pi^2\!-\!9}\,x \,,\qquad x \in [1,\frac{\pi}{3})\,.  
\label{eq:f(x)_X22}
\end{equation}
We now use the general result in Eq.~(\ref{eq:g_i(a)}), and plug in the specific outcomes for $f(x)$, $a(x)$ and $x(a)$ which apply over interval $X_{21}=[\frac{\pi}{6},\frac{\pi}{2}\!-\!1)$. It follows immediately:
\begin{eqnarray}
g_{21}(a) &=& \frac{3}{\pi\!-\!3} \times \left| -\frac{1}{4\sqrt{1-a^2}} \right|\nonumber\\
          &=& \frac{3}{4(\pi\!-\!3)\sqrt{1\!-\!a^2}}\,.
\label{eq:g_21(a)}
\end{eqnarray}
The calculation for $g_{22}(a)$ proceeds as follows. First, we note that in order to obtain inverse $x(a)$ from the equation $a(x)=\cos 4x$ we have multiple possible solutions given by:
\begin{equation}
x(a) = \pm \frac{1}{4}\arccos a + \frac{k\pi}{2}\,,\ \ \ k \in \mathbb{Z}\,.
\end{equation}
However, the fact that $x \in X_{22}=[1,\frac{\pi}{3})$ determines the following solution as the only correct arccos branch in the inverse of $a(x)$: 
\begin{equation}
x(a) = - \frac{1}{4}\arccos a + \frac{\pi}{2}\,.
\end{equation}
It follows immediately from the general result (\ref{eq:g_i(a)}) and the specific outcomes for $f(x)$, $a(x)$ and $x(a)$ which apply in interval $X_{22}=[1,\frac{\pi}{3})$ that:
\begin{eqnarray}
g_{22}(a) &=& \frac{18}{\pi^2 - 9} \left(-\frac{1}{4}\arccos a + \frac{\pi}{2}\right) \times \frac{1}{4\sqrt{1\!-\!a^2}}\nonumber\\
          &=& \frac{9}{8(\pi^2 - 9)}\frac{2\pi-\arccos a}{\sqrt{1\!-\!a^2}}\,.
\label{eq:g_22(a)}
\end{eqnarray}
We remark that by construction both solutions $g_{21}(a)$ in Eq.~(\ref{eq:g_21(a)}) and $g_{22}(a)$ in Eq.~(\ref{eq:g_22(a)}) are normalized to unity over the subset $A_2 = [-0.654,-0.5)$ in question. 

\begin{figure}[h]
\centering
\includegraphics[width=0.5\textwidth]{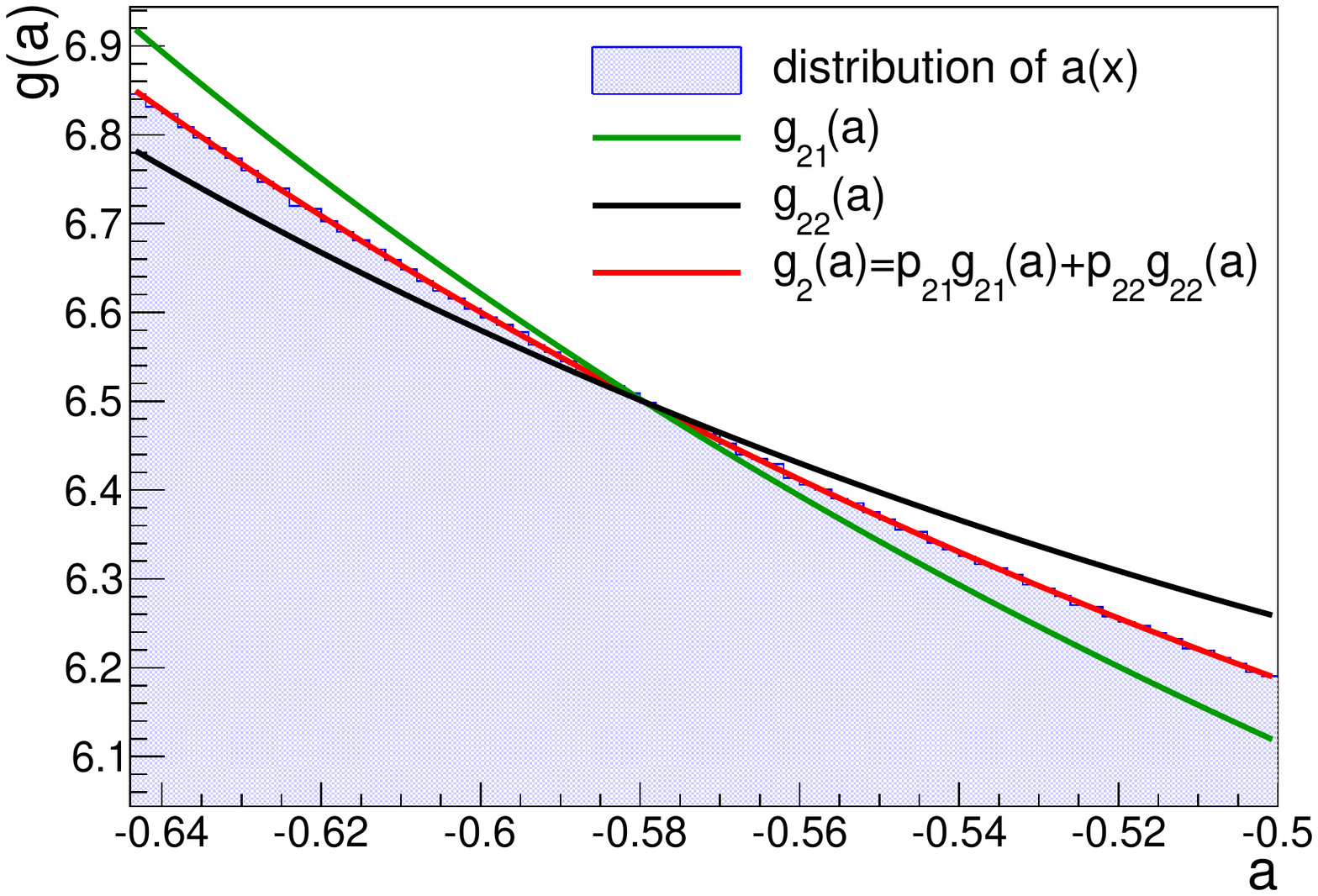}
\caption{(Color online) The resulting distribution of a function $a(x)$ in the subset $A_2$ of a toy Monte Carlo example, together with the derived analytic p.d.f.'s $g(a)$. See the main text for further explanation.}
\label{fig:codomain_A2_0c}
\end{figure}

\begin{figure}[h]
\centering
\includegraphics[width=0.5\textwidth]{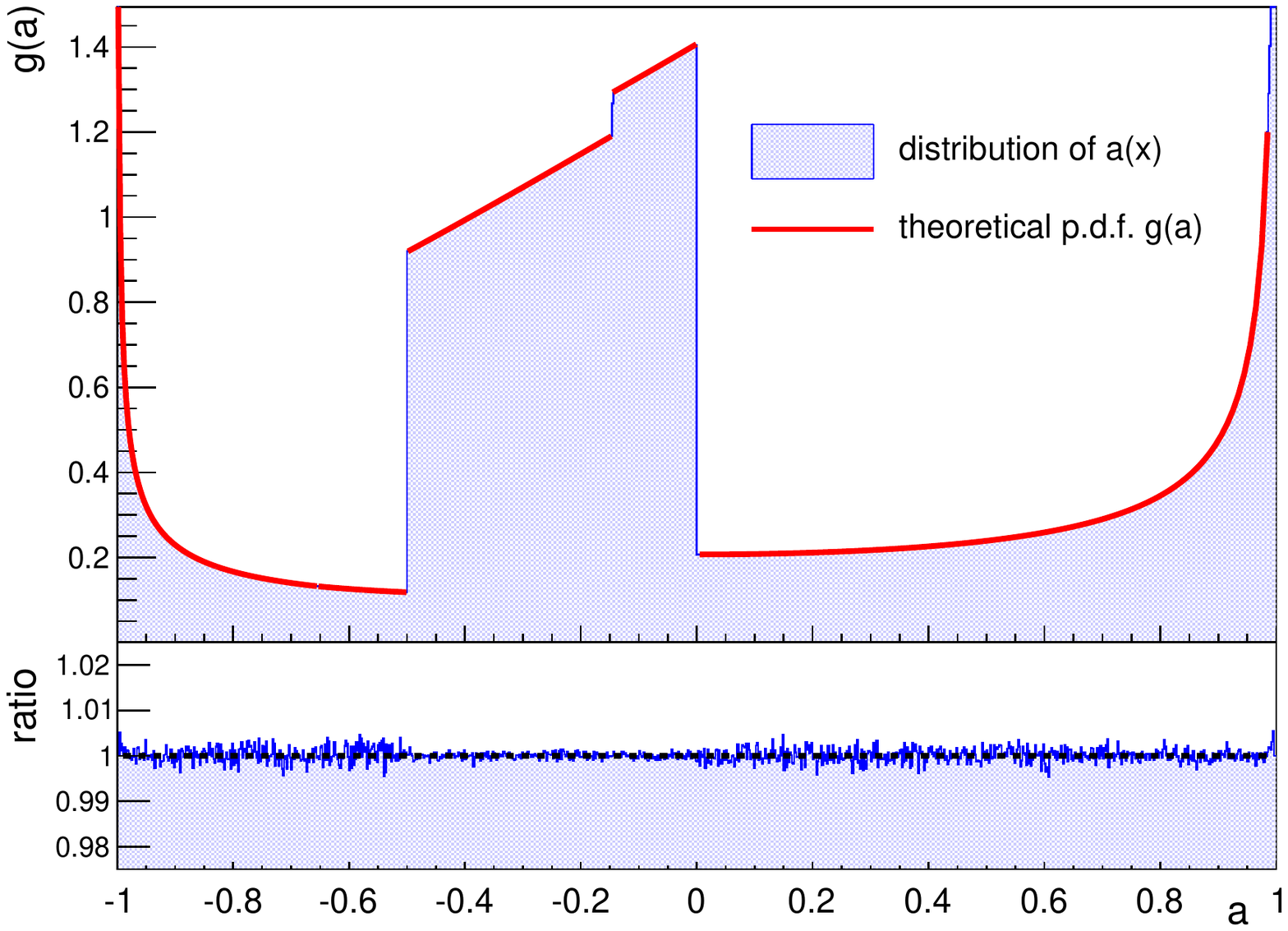}
\caption{(Color online) The resulting distribution (blue histogram) and the derived analytic piecewise p.d.f. $g(a)$ (solid red line) of a function $a(x)$ for the toy Monte Carlo example. On the bottom panel the ratio is shown between resulting distribution and the analytic p.d.f. $g(a)$.}
\label{fig:distribution_of_a_with_pdf_0c}
\end{figure}

Finally, we combine the results (\ref{eq:g_21(a)}) and (\ref{eq:g_22(a)}) by using Eq.~(\ref{eq:g(a)_combined}) in order to obtain the final p.d.f. $g_2(a)$ for subset $A_2$. From Eq.~(\ref{eq:p_ji}) we calculate probabilities $p_{21}$ and $p_{22}$ to obtain straightforwardly:
\begin{eqnarray}
p_{21} &=& \frac{6}{9+\pi}\,,\nonumber\\
p_{22} &=& \frac{3+\pi}{9+\pi}\,.
\label{eq:p21_p22}
\end{eqnarray}
Taking into account results (\ref{eq:g_21(a)}), (\ref{eq:g_22(a)}) and (\ref{eq:p21_p22}), we have obtained by making use of Eq.~(\ref{eq:g(a)_combined}) after some algebra the final solution for the p.d.f. of $a$ in subset $A_2$:
\begin{equation}
g_{2}(a)= \frac{9(4\!+\!2\pi\!-\!\arccos a) }{8(\pi\!+\!9)(\pi\!-\!3)\sqrt{1\!-\!a^2}}\,,\qquad a\in[-0.654,-0.5)\,.
\label{eq:g_2(a)}
\end{equation}
By construction $g_2(a)$ is automatically normalized to unity over $A_2=[-0.654,-0.5)$. We have shown solutions (\ref{eq:g_21(a)}), (\ref{eq:g_22(a)}) and (\ref{eq:g_2(a)}) on Fig.~(\ref{fig:codomain_A2_0c}) in the example in which the sample space of $x$ was restricted only to $X_{21}\cup X_{22}$, i.e. $[\frac{\pi}{6},\frac{\pi}{2}\!-\!1)\cup[1,\frac{\pi}{3})$, so that the total sample space of $a$ is automatically restricted only to $A_2=[-0.654,-0.5)$. With such a setup, the normalized p.d.f. $g_2(a)$ given in (\ref{eq:g_2(a)}) provides exact description of the resulting distribution of $a$ in $A_2$ (see the red line in Fig.~(\ref{fig:codomain_A2_0c})).
On the other hand, if the sample space of $x$ is the whole starting sample space $X=[0,3)$, then the solution (\ref{eq:g_2(a)}) must be scaled with probability $p_2$ that $x$ is mapped with $a(x)$ only into $A_2$. From Eq.~(\ref{eq:p_j}) it follows straightforwardly:
\begin{eqnarray}
p_2 &=&\int_{X_{21}}f(x)\,dx + \int_{X_{22}}f(x)\,dx \nonumber\\
&=& \frac{\pi-3}{15}+\frac{\pi^2-9}{90}\nonumber\\
&=&\frac{1}{90}(\pi+9)(\pi-3)\,.
\label{eq:p2} 
\end{eqnarray}
We now enlist the final results for all five subsets $A_j$ defined in (\ref{eq:Ai's}) of sample space $A$, which were obtained by following the analogous procedure as the one detailed above for $A_2$. We have obtained as our final solution for the p.d.f. $g(a)$ of a function $a(x)$ the following expression:
\begin{equation}
g(a)=
\left\{
 \begin{array}{ll}
 \frac{4-\pi }{10}\!\times\!\frac{1}{(4\!-\!\pi)\sqrt{1\!-\!a^2}}\,,&\qquad a\in [-1,-0.654)\,,\\ 
 \frac{(\pi+9)(\pi-3)}{90}\!\times\!\frac{9(4\!+\!2\pi\!-\!\arccos a) }{8(\pi\!+\!9)(\pi\!-\!3)\sqrt{1\!-\!a^2}}\,,&\qquad a\in[-0.654,-0.5)\,,\\  
 \frac{63\!-\!24\pi\!+\!8\pi^2\!+\!108\cos 8\!+\!18\cos 16}{90}\!\times\!\frac{4\!+\!2\pi\!-\!\arccos a\!+\!32(2a\!+\!3)\sqrt{1\!-\!a^2})}{16\,[\frac{5}{2}\!+\!\frac{4\pi}{9}(\pi\!-\!3)\!+\!2\cos 8(3\!+\!\cos 8)]\sqrt{1\!-\!a^2}}\,,&\qquad a\in[-0.5,-0.146)\,,\\
 \frac{8\!+\!11\pi\!-\!5\pi^2\!-\!48\cos 8\!-\!8\cos 16}{40}\!\times\!\frac{2[1\!+\!\pi\!+8(2a\!+\!3)\sqrt{1\!-\!a^2}\,]}{[(16\!-\!5\pi )(1\!+\!\pi )\!-\!16\cos 8(3\!+\!\cos 8)]\sqrt{1-a^2}}\,,&\qquad a\in[-0.146,0)\,,\\
\frac{\pi(\pi\!+\!1)}{40}\!\times\!\frac{2}{\pi\sqrt{1-a^2}}\,,&\qquad a\in[0,1)\,. 
 \end{array} 
\right.
\label{eq:pdf_g(a)_A}
\end{equation}
The prefactor in each line above denotes the probability that $a(x)\in A_j$, and they were calculated from Eq.~(\ref{eq:p_j}). The analytic piecewise solution for p.d.f. $g(a)$ given in (\ref{eq:pdf_g(a)_A}) is shown with solid red line on Fig.~\ref{fig:distribution_of_a_with_pdf_0c}, together with the actual distribution of $a(x)$ (blue histogram). 
Finally, we remark that each individual analytic solution $g_j(a)$ for p.d.f. of $a$ in the subset $A_j$ can be tested independently by restricting the sample space of $x$ only to those subintervals of $X$ which are mapped with $a(x)$ in $A_j$. In such a setup probability prefactors in (\ref{eq:pdf_g(a)_A}) are not needed, i.e. they are trivially equal to unity, since by construction all sampled $x$ are mapped only into subset $A_j$ under consideration (see again the detailed derivation for subset $A_2$ where such a test has been provided in Fig.~\ref{fig:codomain_A2_0c}). 

\subsection{Few general claims}
\label{app:Few_general_claims}

Now we establish few general claims which, although trivial, simplified a great deal the derivations of our main results. We reuse again the same notation introduced previously in this appendix.  

\vspace{0.44cm}
\noindent\textbf{Claim 1 (``Periodicity"):} If the starting sample space $X$ of a random observable $x$ can be split into $N$ disjoint subsets $X_1, X_2, \ldots,X_N$ for each of which the probability that $x$ will be sampled within it is the same, and if the p.d.f.'s of a function $a(x)$ are all the same for each subset $X_i$ and equal to $g(a)$, then the p.d.f. $g_{\rm union}$ of $a(x)$, which corresponds to the case when the sample space of $x$ is the starting sample space $X$, is given by:  
\begin{equation}
g_{\rm union}(a)=g(a)\,.
\label{eq:Claim_Periodicity}
\end{equation}
\noindent\textit{Proof:} Proof is trivial and it follows solely from relation (\ref{eq:g(a)_combined}) and observation that probabilities are normalized under the assumptions used in the setup of this claim. This claim in particular clarifies that for periodic functions $a(x)$ the p.d.f. $g(a)$ obtained only for one fundamental interval of $a(x)$ is exactly the same as the p.d.f. $g(a)$ for the whole range over which $a(x)$ is defined.    \hspace*{\fill} $\square$ 

\vspace{0.44cm}
\noindent\textbf{Claim 2 (``Signature"):} If the p.d.f. $g(a)$ of random observable $a$ is known and if $b\equiv-a$, than the p.d.f. of $b$, denoted by $h(b)$, is given by: 
\begin{equation}
h(b) = g(-b)\,.
\label{eq:Claim_Signature}
\end{equation}

\noindent\textit{Proof:} Proof is trivial and it follows immediately from Eq.~(\ref{eq:g_i(a)}) which gives in this context: 
\begin{eqnarray}
h(b) &=& g(a(b))\left|\frac{da}{db}\right|\nonumber\\
 &=& g(-b)|-1|\nonumber\\
 &=& g(-b)\,.   
\end{eqnarray}
As an example, in practice it suffices to obtain p.d.f. of $\sin mx$ only for positive integers $m$, which then can be trivially extended also to negative values of $m$ by using the outcome of this claim. \hspace*{\fill}$\square$


\section{Fundamental results}
\label{app:Fundamental_results}


\noindent In this appendix we provide the detailed derivation of all fundamental results which were used in the derivation of more complex cases. All results presented here were obtained from the general procedure established in the Appendix~\ref{app:Distributions_of_functions_of_random_observables}, which therefore shall be read first. In this appendix we again largely follow the notation utilized in classical textbook~\cite{Cowan:1998ji}. The starting random observable we denote by $x$, its sample space by $X$ and its p.d.f. by $f(x)$. The function of $x$ we denote by $a(x)$, which itself is a random observable with sample space $A$ and p.d.f. $g(a)$. We select $a(x)$ to be either $\cos mx$ or $\sin mx$, where $m$ is a non-zero integer, and we derive analytically its p.d.f. $g(a)$ for the following specific choices of interest for $f(x)$: a) random walk; and b) monochromatic flow. Although in general we allow $m$ to be a nonzero integer, we will focus during derivations only on positive integers $m$ and trivially extend the final results to negative integers $m$ only at the end of calculation by using the outcome of claim~(\ref{eq:Claim_Signature}).  

\subsection{Random walk}
\label{app:Random_walk}

We start with the simplest problem in which observable $x$ is sampled randomly, and we derive the p.d.f. $g(a)$ of its function $a(x)$ for two concrete cases of interest. 

\vspace{0.44cm}
\noindent{\textbf{Case 1:}} $x\in [0,2\pi)$, $f(x)=\frac{1}{2\pi}$, $a(x)=\cos mx$.
\vspace{0.44cm}

\noindent In this case $a(x)$ does not have a unique inverse over the whole interval $[0,2\pi)$. Instead, we have $2m$ disjoint subsets, where the boundaries of the $i$th subset over which $a(x)$ has a unique inverse are given by $[\frac{(i-1)\pi}{m},\frac{i\pi}{m}), i = 1, 2,\ldots,2m$. The normalized p.d.f. $f_i(x)$ in the $i$th subset $[\frac{(i-1)\pi}{m},\frac{i\pi}{m})$ is given by:
\begin{equation}
f_i(x) = \frac{m}{\pi}\,.
\label{eq:f_i_RW_Case_1}
\end{equation}
We use Eq.~(\ref{eq:g_i(a)}) which yields for the $i$th subset:
\begin{eqnarray}
g_i(a) &=& \frac{m}{\pi}\cdot\left|\frac{1}{m}(-\frac{1}{\sqrt{1-a^2}})\right|\nonumber\\
&=&\frac{1}{\pi}\frac{1}{\sqrt{1-a^2}}\,.
\end{eqnarray}
The probability $p_i$ that $x$ is sampled in the $i$th subset is equal to $\frac{1}{2m}$ for any $i$. Therefore, we can apply the claim (\ref{eq:Claim_Periodicity}) to obtain immediately our final result:
\begin{equation}
g(a)=\frac{1}{\pi}\frac{1}{\sqrt{1-a^2}}\,,\qquad a\in[-1,1)\,.
\label{eq:RW_cos}
\end{equation}
We see that the p.d.f. $g(a)$ of a function $a(x)=\cos mx$, when $x$ itself is sampled randomly, is exactly the same function for all choices of non-zero integers $m$.

\vspace{0.44cm}
\noindent{\textbf{Case 2:}} $x\in [0,2\pi)$, $f(x)=\frac{1}{2\pi}$, $a(x)=\sin mx$.
\vspace{0.44cm}

\noindent Since $a(x)$ does not have a unique inverse over the whole interval $x\in [0,2\pi)$, we fragment it into disjoint intervals with boundaries $[0,\frac{\pi}{2m})$, $[\frac{\pi}{m}(i+\frac{1}{2}),\frac{\pi}{m}(i+\frac{3}{2})), i = 0,1, 2,\ldots,2m-2$, and $[\frac{\pi}{m}(2m-\frac{1}{2}),2\pi)$. We map the boundaries of all these subsets with $a(x)$ to obtain set of values $-1,0,1$, from which we read off the two disjoint subsets of $A$, namely $A_1=[-1,0)$ and $A_2=[0,1)$. We work out the results in these two subsets independently. 

\noindent $A_1$: There are $2m$ intervals which contribute to this part of codomain.  The normalized p.d.f. $f(x)$ in the first interval $X_{11}=[\frac{\pi}{m},\frac{3\pi}{2m})$ is trivially $f_{11}(x)=\frac{2m}{\pi}$. Therefore:
\begin{eqnarray}
g_{11}&=&\frac{2m}{\pi}\cdot \left|\frac{1}{m\sqrt{1-a^2}}\right|\nonumber\\
&=& \frac{2}{\pi\sqrt{1-a^2}}\,.
\end{eqnarray}
Since the calculation for all other intervals are exactly the same, the claim (\ref{eq:Claim_Periodicity}) applies so  we have immediately our final result for $A_1$:
\begin{equation}
g_1(a) = \frac{2}{\pi\sqrt{1-a^2}}\,,\qquad a \in [-1,0)\,.
\label{eq:RW_case2_g_1}
\end{equation}

\noindent $A_2$: Again, there are $2m$ intervals which contribute. The normalized p.d.f. $f(x)$ in the first interval $X_{21}=[0,\frac{\pi}{2m})$ is trivially $f_{21}(x)=\frac{2m}{\pi}$. The rest of the calculation is exactly the same as for $A_1$ above, so we just write the final result for $A_2$:
\begin{equation}
g_2(a) = \frac{2}{\pi\sqrt{1-a^2}}\,,\qquad a \in [0,1)\,.
\label{eq:RW_case2_g_2}
\end{equation}
Given the solutions (\ref{eq:RW_case2_g_1}) and (\ref{eq:RW_case2_g_2}) above, we can trivially extend them over all codomain (with trivial different normalization due to different length of the sample space), to obtain our final result for $A$:  
\begin{equation}
g(a) = \frac{1}{\pi\sqrt{1-a^2}}\,,\qquad a\in [-1,1)\,.
\label{eq:RW_sin}
\end{equation}
Again as in the previous case, the p.d.f. $g(a)$ of a function $a(x)=\sin mx$, when $x$ is sampled randomly, is exactly the same for all choices of non-zero integers $m$. We now move on to the case of monochromatic flow. 

\subsection{Monochromatic flow}
\label{app:Monochromatic_flow}

Now we tackle a problem in which observable $x$ is sampled from a Fourier-like p.d.f. $f(x)$ parametrized only with one degree of freedom, and we are interested in finding p.d.f. $g(a)$ of the function $a(x)$ for two cases of interest, namely $a(x)=\cos mx$ and $a(x)=\sin mx$, where $m$ is a nonzero integer. One degree of freedom in Fourier series can be associated either with cosine or sinus part, so in total we have four distinct cases to consider here. We will frequently in the derivations in this section use orthogonality relations of trigonometric functions, which therefore for convenience sake we enlist now:
\begin{eqnarray}
\int_{0}^{2\pi}\!\sin(mx\!+\!\alpha)\sin(nx\!+\!\beta)dx &=& \pi\cos(\alpha\!-\!\beta)\delta_{mn}\,,\label{eq:sin_sin}\\
\int_{0}^{2\pi}\!\cos(mx\!+\!\alpha)\cos(nx\!+\!\beta)dx &=& \pi\cos(\alpha\!-\!\beta)\delta_{mn}\,,\label{eq:cos_cos}\\
\int_{0}^{2\pi}\!\sin(mx\!+\!\alpha)\cos(nx\!+\!\beta)dx &=& \pi\sin(\alpha\!-\!\beta)\delta_{mn}\,,\label{eq:sin_cos}
\end{eqnarray}
where $\delta_{mn}$ is the Kronecker delta symbol, $n$ and $m$ are nonzero integers, and $\alpha$ and $\beta$ are arbitrary. For the same reason, we also enlist the following relations to which we will frequently refer to:
\begin{eqnarray}
\cos y = x &\Leftrightarrow& y = \pm\arccos x + 2k\pi\,, \label{eq:arccos_branches} \\
\sin y = x &\Leftrightarrow& y = (-1)^k\arcsin x + k\pi\,, \label{eq:arcsin_branches} 
\end{eqnarray}
where $k$ is some integer.
 
\vspace{0.44cm}
\noindent{\textbf{Case 3:}} $x\in [0,2\pi)$, $f(x)=\frac{1}{2\pi}(1+2c_n\cos nx)$, $a(x)=\cos mx$.
\vspace{0.44cm}

\begin{figure}[h]
\centering
\includegraphics[width=0.5\textwidth]{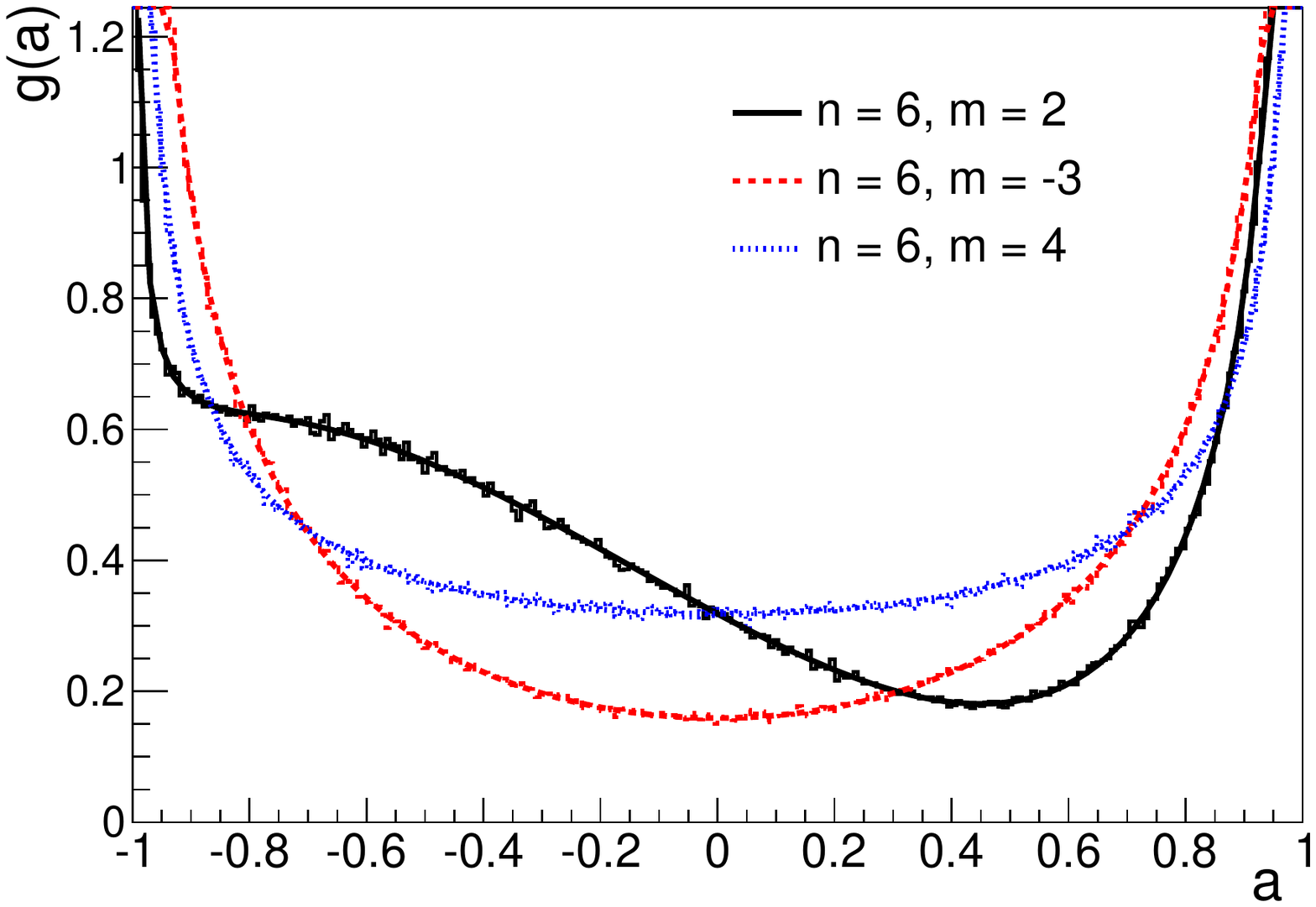}
\caption{(Color online) The resulting distributions of a function $a(x) = \cos mx$ in a toy Monte Carlo example for monochromatic flow parametrized only with harmonic $c_6 = 0.25$ and analytic results from Eq.~(\ref{eq:g(a)_Cos_Cos}), for $m=2$ (black), $m=-3$ (red) and $m=4$ (blue).}
\label{fig:Case_3_Cos_Cos_0c}
\end{figure}

\noindent We first observe that $a(x)$ does not have a unique inverse over the whole interval $[0,2\pi)$, but instead there are $2m$ disjoint subsets of $X$ in each of which $a(x)$ has a unique inverse. The boundaries of the $i$th such subset are given by $[\frac{(i-1)\pi}{m},\frac{i\pi}{m}), i = 1, 2,\ldots,2m$. All boundaries are mapped with $a(x)$ into either -1 or 1, from which we conclude that for all $2m$ subsets the codomain of $a(x)$ is the same and it equals to the whole sample space $A = [-1,1)$. Next, we show that only the cases in which $n/m$ is an {\it arbitrary} integer yield to solutions for $g(a)$ which are not the same as the solution for the random walk given in Eq.~(\ref{eq:RW_cos}). To establish this claim, we start from the following equality~\cite{Cowan:1998ji}:
\begin{equation}
\int a^p g(a) da = \int a^{p}(x) f(x) dx\,,
\label{eq:integral_equality_from_Cowan}
\end{equation}
where $p$ is some positive integer. If $p$ is even, than we can write $p\equiv 2r$, and use the relation (1.320.5) from~\cite{GR:2007} to obtain:
\begin{equation}
\cos^{2r}mx=\frac{1}{2^{2r}}\left[\sum_{k=0}^{r-1}2\binom{2r}{k}\cos 2(r-k)mx+\binom{2r}{r}\right]\,.
\label{eq:RS_cos^2r}
\end{equation}
Using relations (\ref{eq:RS_cos^2r}) and (\ref{eq:cos_cos}), we see immediately that due to the presence of Kronecker symbol in (\ref{eq:cos_cos}), the integral on the RHS in (\ref{eq:integral_equality_from_Cowan}) will have only contributions from the terms  for which $2(r-k)m = n$ is satisfied, i.e. 
\begin{equation}
2(r-k) = \frac{n}{m}\,.
\end{equation}
Since on the LHS is an even integer expression, the above equality can hold only for those integers $n$ and $m$ for which their ratio is also an even integer, which partially establishes our starting claim. In order to complete it, 
we also have to consider the case when $p$ in Eq.~(\ref{eq:integral_equality_from_Cowan}) is odd, i.e. $p\equiv 2r-1$. We can use the relation (1.320.7) from~\cite{GR:2007} to obtain:
\begin{equation}
\cos^{2r-1}mx=\frac{1}{2^{2r-2}}\,\sum_{k=0}^{r-1}\binom{2r-1}{k}\cos (2r\!-\!2k\!-\!1)mx\,.
\label{eq:RS_cos^2r-1}
\end{equation}
By following the same reasoning as in the previous case in which $p$ was even, we obtain that the integral on RHS in (\ref{eq:integral_equality_from_Cowan}) will have only contributions from the terms for which $(2r\!-\!2k\!-\!1)m = n$ is satisfied, i.e.
\begin{equation}
2r\!-\!2k\!-\!1 = \frac{n}{m}\,.
\end{equation}
Since on the LHS is an odd integer, the above equality will hold only for those integers $n$ and $m$ for which their ratio is also an odd integer, which completes our starting claim. Using these results, the normalized p.d.f. $f_i(x)$ in the $i$th interval $[\frac{(i-1)\pi}{m},\frac{i\pi}{m})$ is given by:
\begin{equation}
f_i(x)=\frac{m}{\pi}(1+2c_n\cos nx)\,, \qquad  n/m{\rm\ is\ an\ arbitrary\ integer}  \,.
\end{equation}
Therefore, from Eq.~(\ref{eq:g_i(a)}) we obtain that in the $i$th interval:
\begin{equation}
g_i(a) = \frac{m}{\pi}\bigg[1+2c_n \cos\big(\frac{n}{m}\arccos a\big)\bigg]\left|\frac{1}{m}\frac{-1}{\sqrt{1-a^2}}\right|\,, \qquad  n/m{\rm\ is\ an\ arbitrary\ integer} \,.
\label{eq:temp20140828_0}
\end{equation}
To make a further progress, we recall the trigonometric definition of $n$th Chebyshev polynomial of the first kind:
\begin{equation}
T_n(x) \equiv \cos(n\arccos x)\,,
\label{eq:Chebyshev_polynomial_first_kind_trigonometric_definition}
\end{equation}
which we can use directly in our case due to the fact that $n/m$ is an integer. It follows immediately from (\ref{eq:temp20140828_0}) that for the $i$th interval:
\begin{equation}
g_i(a) = \frac{1+2c_nT_{\frac{n}{m}}(a)}{\pi\sqrt{1-a^2}}\,, \qquad n/m{\rm\ is\ an\ arbitrary\ integer} \,.
\end{equation}
Since the conditions from claim (\ref{eq:Claim_Periodicity}) apply in this case, we can immediately write our final analytic solution:
\begin{equation}
g(a)=
\left\{
 \begin{array}{ll}
  \frac{1+2c_nT_{\frac{n}{m}}(a)}{\pi\sqrt{1-a^2}}\,,& n/m{\rm\ is\ an\ arbitrary\ integer}\,, \\
  \frac{1}{\pi\sqrt{1-a^2}}\,,& {\rm otherwise}\,, \\  
 \end{array} 
\right.
\label{eq:g(a)_Cos_Cos}
\end{equation}
where $a\in[-1,1)$ and $T_n$ is $n$th Chebyshev polynomial of the first kind. Although the above derivation was carried out explicitly only for positive integers $m$, we can extend trivially due to the fact that $\cos mx = \cos |m|x$ the solution (\ref{eq:g(a)_Cos_Cos}) to negative integers $m$ by replacing $n/m$ with $n/|m|$ in the index of Chebyshev polynomial. We have tested the analytic solution (\ref{eq:g(a)_Cos_Cos}) in a simple Monte Carlo study presented on Fig.~\ref{fig:Case_3_Cos_Cos_0c}, for $n=6$, $c_6 = 0.25$ and $m=2,-3,4$.  
%


\vspace{0.44cm}
\noindent{\textbf{Case 4:}} $x\in [0,2\pi)$, $f(x)=\frac{1}{2\pi}(1+2c_n\cos nx)$, $a(x)=\sin mx$.
\vspace{0.44cm}

\noindent Since $a(x)$ does not have a unique inverse over the whole interval $x\in [0,2\pi)$, we fragment it into disjoint subintervals with boundaries $[0,\frac{\pi}{2m})$, $[\frac{\pi}{m}(i\!+\!\frac{1}{2}),\frac{\pi}{m}(i\!+\!\frac{3}{2})), i = 0,1, 2,\ldots,2m-2$, and $[\frac{\pi}{m}(2m\!-\!\frac{1}{2}),2\pi)$. The first and last subinterval are mapped with $a(x)$ into codomains $A_f=[0,1)$ and $A_l[-1,0)$, respectively, while all remaining subintervals are for any $i$ mapped into the whole sample space $A=[-1,1)$. Before proceeding further, we demonstrate that only the cases in which $n/m$ is an {\it even} integer yield to solutions for $g(a)$ which are not the same as the random walk solution in Eq.~(\ref{eq:RW_sin}). In order to prove this claim, we again focus on equality~(\ref{eq:integral_equality_from_Cowan}), and we consider cases when $p$ is even or odd integer separately. For the case when $p$ is even, we can write $p\equiv 2r$ and we use the relation (1.320.1) from~\cite{GR:2007} to obtain:
\begin{equation}
\sin^{2r}mx=\frac{1}{2^{2r}}\left[\sum_{k=0}^{r-1}(-1)^{r-k}\,2\binom{2r}{k}\cos 2(r-k)mx+\binom{2r}{r}\right]\,.
\label{eq:RS_sin^2r}
\end{equation}
Using relations (\ref{eq:RS_sin^2r}) and (\ref{eq:cos_cos}), we see immediately that due to the presence of Kronecker symbol in (\ref{eq:cos_cos}) the integral on the RHS in (\ref{eq:integral_equality_from_Cowan}) will have only contributions from the terms  for which $2(r-k)m = n$ is satisfied, i.e. 
\begin{equation}
2(r-k) = \frac{n}{m}\,.
\end{equation}
Since on the LHS is an even integer, the above equality can hold only for those integers $n$ and $m$ for which their ratio is also an even integer. On the other hand, when $p$ in Eq.~(\ref{eq:integral_equality_from_Cowan}) is odd, i.e. $p\equiv 2r-1$, we can use the relation (1.320.3) from~\cite{GR:2007} to obtain:
\begin{equation}
\sin^{2r-1}mx=\frac{1}{2^{2r-2}}\,\sum_{k=0}^{r-1}(-1)^{r+k-1}\binom{2r-1}{k}\sin (2r\!-\!2k-\!1)mx\,.
\label{eq:RS_sin^2r-1}
\end{equation}
The relevant orthogonality relation in this case is (\ref{eq:sin_cos}) with $\alpha = \beta = 0$, which always yields zero irrespectively of the value of Kronecker symbol, due to the presence of $\sin(\alpha\!-\!\beta)$ term. Therefore, when $p$ is odd there is no a single term which can contribute, which completes our claim. We now proceed with calculation by considering only the cases in which the ratio $n/m$ is an even integer.

\begin{figure}[h]
\centering
\includegraphics[width=0.5\textwidth]{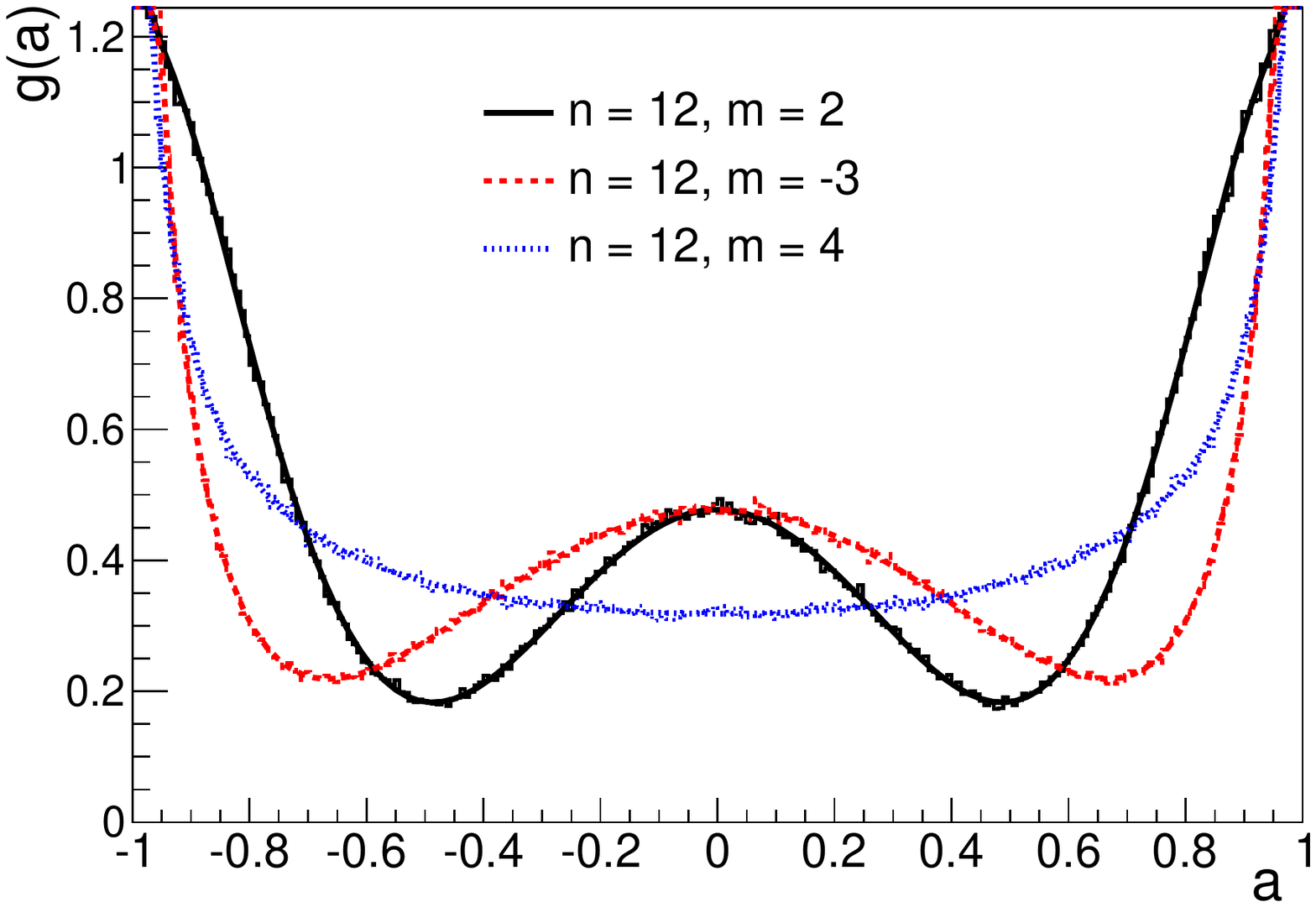}
\caption{(Color online) The resulting distributions of a function $a(x) = \sin mx$ in a toy Monte Carlo example for monochromatic flow parametrized only with harmonic $c_{12} = 0.25$ and analytic results from Eq.~(\ref{eq:g(a)_Cos_Sin}), for $m=2$ (black), $m=-3$ (red) and $m=4$ (blue).}
\label{fig:Case_4_Cos_Sin_0a}
\end{figure}

We start by working out the solution for the $i$th interval with boundaries $[\frac{\pi}{m}(i\!+\!\frac{1}{2}),\frac{\pi}{m}(i\!+\!\frac{3}{2})), i = 0,1, 2,\ldots,2m-2$, and which is mapped with $a(x)$ into the whole codomain $A=[-1,1)$ for any $i$. When calculating the inverse of $\sin mx$ in the $i$th interval, a special care has to be taken that the correct $\arcsin$ branch is selected which applies over that interval. From (\ref{eq:arcsin_branches}) and from the parametrization of the boundaries, we obtain that the correct $\arcsin$ branch for the inverse $x(a)$ in the $i$th interval is given by:
\begin{equation}
a(x)=\sin mx \Leftrightarrow x(a)=\frac{(-1)^{i+1}}{m}\arcsin a + \frac{(i+1)\pi}{m}\,. 
\end{equation}
It follows:
\begin{eqnarray}
\cos n x(a) &=& \cos\left[\frac{n}{m}(-1)^{i+1}\arcsin a + \frac{n}{m}(i+1)\pi\right]\nonumber\\
&=&\cos\left(\frac{n}{m}\arcsin a\right)\,.
\end{eqnarray}
In order to obtain the last equality above, we have used the fact that cosine is an even function, and that in the case under consideration $n/m$ is an even integer (if $n/m$ was an arbitrary integer, than the result after last equality above has to be multiplied with prefactor $(-1)^{n(i+1)/m}$, which always evaluates to 1 when $n/m$ is an even integer). In order to make further progress, we observe that from the trigonometric definition (\ref{eq:Chebyshev_polynomial_first_kind_trigonometric_definition}) of Chebyshev polynomials of the first kind we have: 
\begin{equation}
\cos(n\arcsin a) = T_n(\cos\arcsin a) = T_n(\sqrt{1-a^2})\,.
\end{equation}
Combining the expression above with the alternative definition of Chebyshev's polynomial of the first kind given as relation (8.940.1) in~\cite{GR:2007}:
\begin{equation}
T_n(x) = \frac{1}{2}\left[(x+i\sqrt{1-x^2})^n+(x-i\sqrt{1-x^2})^n\right]\,,
\end{equation}
we obtain after some straightforward algebra the following relation which holds when $n$ is even integer:
\begin{equation}
T_n(\sqrt{1-x^2}) = i^n\,T_n(x)\,,\qquad  n {\rm\ is\ an\ even\ integer}\,.
\end{equation}
Under the working condition that $n/m$ is even integer, the normalized p.d.f. $f_i(x)$ in the $i$th interval is:
\begin{equation}
f_i(x)=\frac{m}{\pi}(1+2c_n\cos nx)\,,\qquad n/m {\rm\ is\ an\ even\ integer}\,.
\end{equation}
Combining all above results together, we obtain from Eq.~(\ref{eq:g_i(a)}) that for the $i$th interval the solution is (note the different meaning of symbol $i$ on two sides of equality):
\begin{equation}
g_i(a)=
\left\{
 \begin{array}{ll}
  \frac{1+2c_ni^{\frac{n}{m}}T_{\frac{n}{m}}(a)}{\pi\sqrt{1-a^2}}\,,& n/m\ {\rm is\ an\ even\ integer}\,, \\
  \frac{1}{\pi\sqrt{1-a^2}}\,,& {\rm otherwise}\,. \\  
 \end{array} 
\right.
\label{eq:gi_temp_20140830}
\end{equation}
We now solve the two remaining cases which are mapped in different codomains, namely for the first subinterval $[0,\frac{\pi}{2m})$ and for the last one with boundaries $[\frac{\pi}{m}(2m\!-\!\frac{1}{2}),2\pi)$, with codomains $A_f = [0,1)$ and $A_l = [-1,0)$, respectively. Trivially, if we make a union of these two subintervals, this problem is completely equivalent to the previous cases, and therefore for the union of these two subintervals we have obtained the same solution as in (\ref{eq:gi_temp_20140830}). Therefore we can again use the result (\ref{eq:Claim_Periodicity}) to write immediately our final analytic solution:
\begin{equation}
g(a)=
\left\{
 \begin{array}{ll}
  \frac{1+2c_ni^{\frac{n}{m}}T_{\frac{n}{m}}(a)}{\pi\sqrt{1-a^2}}\,,& n/m\ {\rm is\ an\ even\ integer}\,, \\
  \frac{1}{\pi\sqrt{1-a^2}}\,,& {\rm otherwise}\,, \\  
 \end{array} 
\right.
\label{eq:g(a)_Cos_Sin} 
\end{equation}
where $a\in[-1,1)$. The above derivation was carried out for positive integers $m$, but it can be easily concluded that the result (\ref{eq:g(a)_Cos_Sin}) applies for negative integers $m$ as well, after replacing $m$ with $|m|$ everywhere in Eq.~(\ref{eq:g(a)_Cos_Sin}). In particular, this conclusion follows trivially from claim~(\ref{eq:Claim_Signature}) and the fact that $n/|m|$ is even integer in the case under consideration here, when the Chebyshev polynomials of the first kind are even functions, i.e. $T_n(a) = T_n(-a)$ if $n$ is an even integer. We also remark on the important role of factor $i^{n/m}$ in (\ref{eq:g(a)_Cos_Sin}), which evaluates to 1 or -1 depending on whether an even integer $n/m$ in addition satisfies also the relation $n/m\ {\rm mod\ 4} = 0$ (when it evaluates to 1) or not (when it evaluates to -1). We have tested the analytic solution (\ref{eq:g(a)_Cos_Sin}) in a simple Monte Carlo study presented on Fig.~\ref{fig:Case_4_Cos_Sin_0a}, for $n=12$, $c_{12} = 0.25$ and $m=2,-3,4$.  
%


\vspace{0.44cm}
\noindent{\textbf{Case 5:}} $x\in [0,2\pi)$, $f(x)=\frac{1}{2\pi}(1+2s_n\sin nx)$, $a(x)=\cos mx$.
\vspace{0.44cm}

\begin{figure}[h]
\centering
\includegraphics[width=0.5\textwidth]{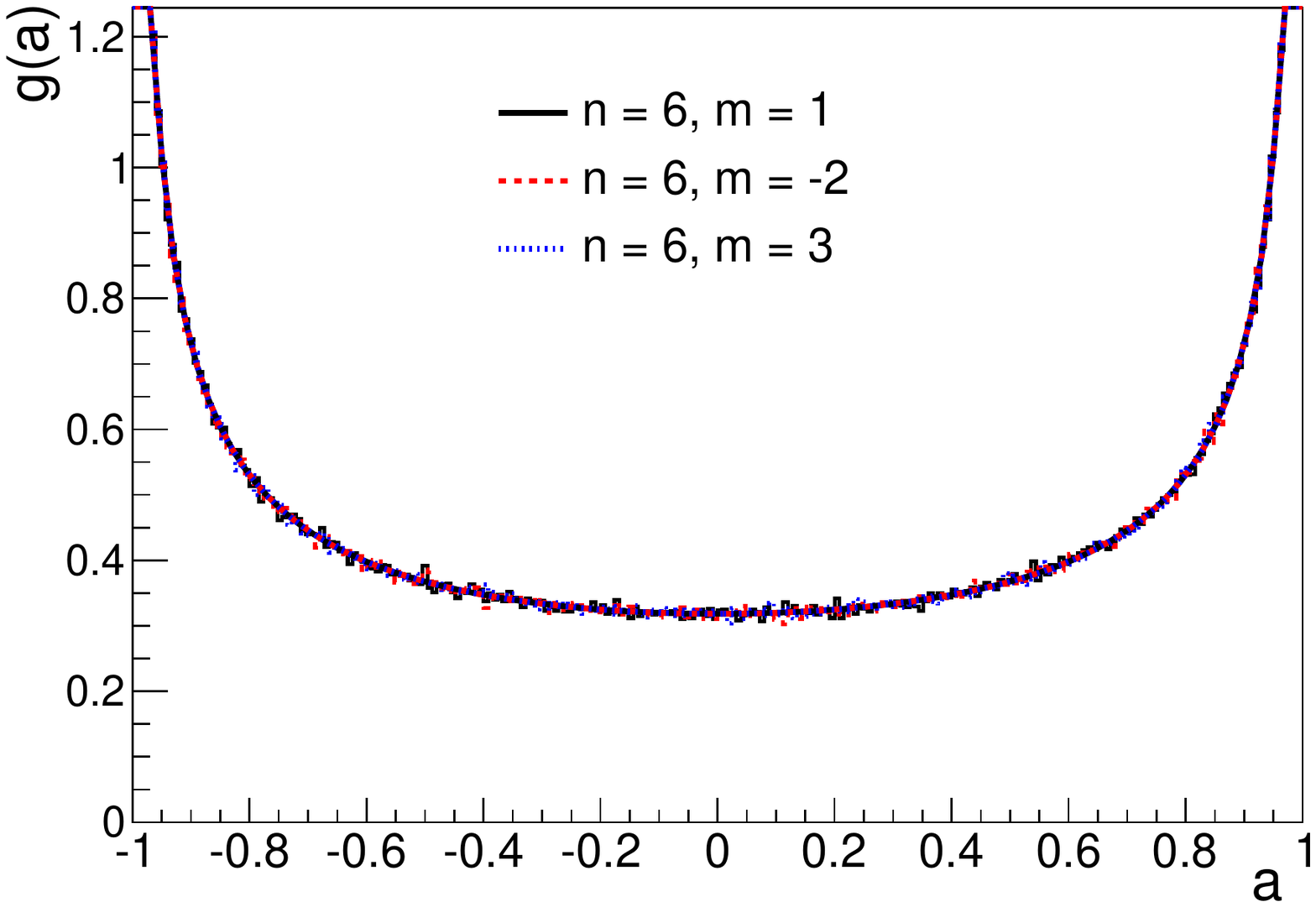}
\caption{(Color online) The resulting distributions of a function $a(x) = \cos mx$ in a toy Monte Carlo example for monochromatic flow parametrized only with harmonic $s_{6} = 0.25$ and analytic results from Eq.~(\ref{eq:g(a)_Sin_Cos}), for $m=1$ (black), $m=-2$ (red) and $m=3$ (blue).}
\label{fig:Case_5_Sin_Cos_0a}
\end{figure}

\noindent This is the simplest case. For any choice of integers $n$ and $m$ we have obtained the same solution as for the random walk, namely: 
\begin{equation}
g(a)=\frac{1}{\pi\sqrt{1-a^2}}\,,
\label{eq:g(a)_Sin_Cos}
\end{equation}
where $a\in [-1,1)$. This can be immediately concluded from Eq.~(\ref{eq:integral_equality_from_Cowan}) by following the same reasoning as in Case 3 and in Case 4, and by observing that only cosine terms appear in both decompositions (\ref{eq:RS_cos^2r}) and (\ref{eq:RS_cos^2r-1}), all of which then in combination with orthogonality relation (\ref{eq:sin_cos}) yields zero contribution to $g(a)$, due to the fact that $\alpha = \beta = 0$ in our case, and the presence of $\sin(\alpha\!-\!\beta)$ term in orthogonality relation (\ref{eq:sin_cos}). We have tested the analytic solution (\ref{eq:g(a)_Sin_Cos}) in a simple Monte Carlo study presented on Fig.~\ref{fig:Case_5_Sin_Cos_0a}, for $n=6$, $s_6 = 0.25$ and $m=1,-2,3$.  
%


\vspace{0.44cm}
\noindent{\textbf{Case 6:}} $x\in [0,2\pi)$, $f(x)=\frac{1}{2\pi}(1+2s_n\sin nx)$, $a(x)=\sin mx$.
\vspace{0.44cm}

\begin{figure}[h]
\centering
\includegraphics[width=0.5\textwidth]{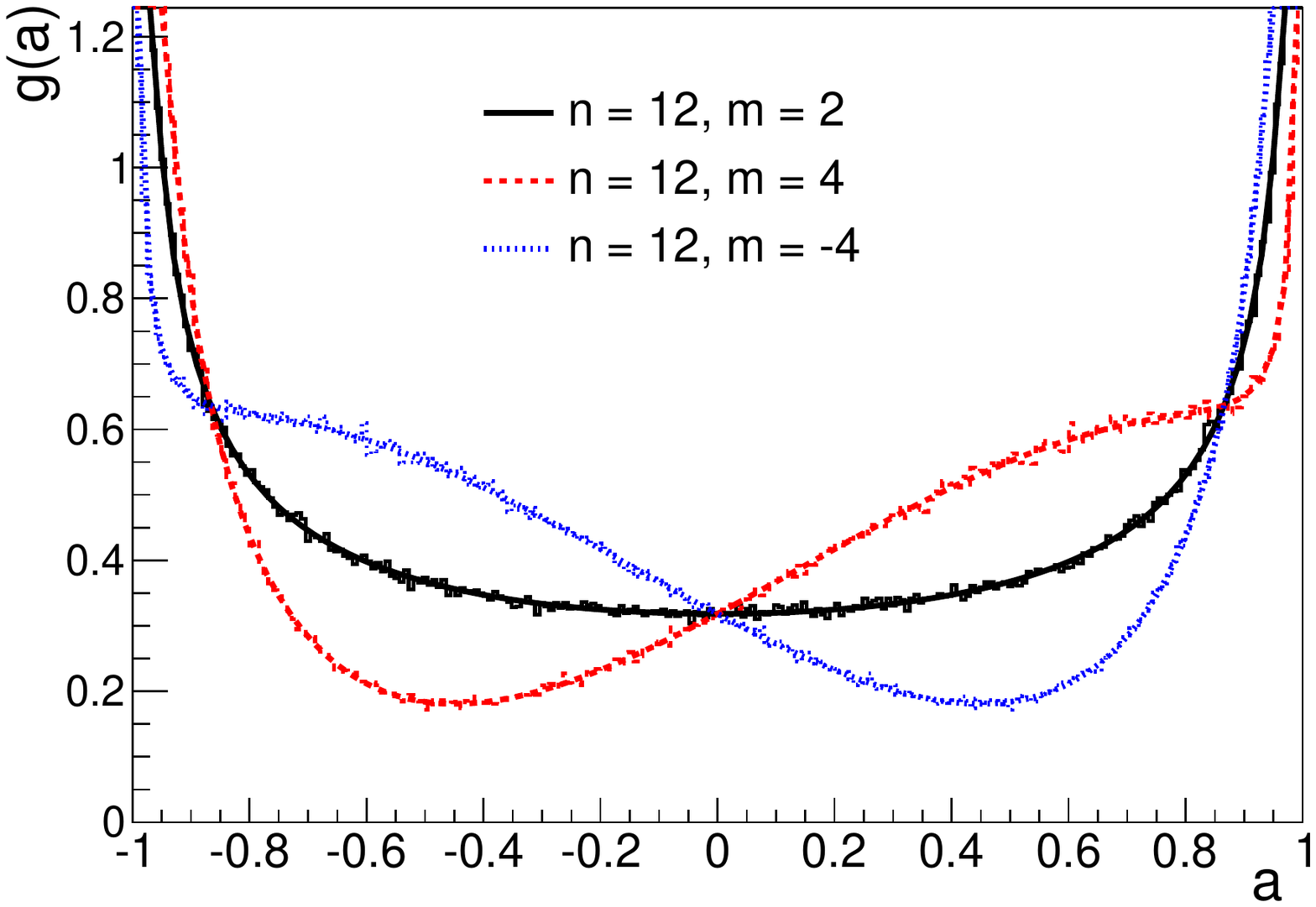}
\caption{(Color online) The resulting distributions of a function $a(x) = \sin mx$ in a toy Monte Carlo example for monochromatic flow parametrized only with harmonic $s_{12} = 0.25$ and analytic results from Eq.~(\ref{eq:g(a)_Sin_Sin}) (see also the discussion below it which concerns the negative integers $m$), for $m=2$ (black), $m=4$ (red) and $m=-4$ (blue).}
\label{fig:Case_6_Sin_Sin_0a}
\end{figure}

\noindent In this last case we outline majority of the statements without proofs, because the details are the same as in Case 3 and Case 4 detailed before. Now only when $n/m$ is an {\it odd} integer the solutions for $g(a)$ are not the same as the random walk solution in Eq.~(\ref{eq:RW_sin}). In order to find an expression for $\sin (n\arcsin x)$, where $n$ is an integer, we start from the definition of Chebyshev's polynomials of the second kind given as relation (8.940.2) in~\cite{GR:2007}:
\begin{equation}
U_n(x) \equiv \frac{\sin[(n+1)\arccos x]}{\sin\arccos x}\,.
\end{equation}
We use the above equality for the case $n\rightarrow n-1$ and $x\rightarrow \cos\arcsin a = \sqrt{1-a^2}$ to obtain:
\begin{equation}
\sin (n\arcsin a) = a\,U_{n-1}(\sqrt{1-a^2})\,.
\end{equation}
With some straightforward algebra, one can show that for an odd $n$ we have the following relation between Chebyshev's polynomials of the first and second kind:
\begin{equation}
U_{n-1}(\sqrt{1-a^2}) = \frac{i^{n-1}}{a}\,T_n(a)\,,\qquad n\ {\rm is\ an\ odd\ integer}\,.
\end{equation}
Therefore, 
\begin{equation}
\sin (n\arcsin a) = i^{n-1}\,T_n(a)\,,\qquad n\ {\rm is\ an\ odd\ integer}\,.
\end{equation}
The rest of calculation is completely analogous as in previous cases. Our final solution is:
\begin{equation}
g(a)=
\left\{
 \begin{array}{ll}
  \frac{1+2s_n\,i^{\frac{n}{m}-1}T_{\frac{n}{m}}(a)}{\pi\sqrt{1-a^2}}\,,& n/m\ {\rm is\ an\ odd\ integer}\,, \\
  \frac{1}{\pi\sqrt{1-a^2}}\,,& {\rm otherwise}\,, \\  
 \end{array} 
\right.
\label{eq:g(a)_Sin_Sin}
\end{equation}
where $a\in[-1,1)$. The above derivation was carried out for positive integers $m$, but the final solution (\ref{eq:g(a)_Sin_Sin}) can be easily extended for negative integers $m$ as well. Namely, one has simply to replace $i^{\frac{n}{m}-1}T_{\frac{n}{m}}$ with ${\rm sgn}(m)i^{\frac{n}{|m|}-1}T_{\frac{n}{|m|}}$ in (\ref{eq:g(a)_Sin_Sin}), and then this result applies both for positive and negative integers $m$. This conclusion follows simply from claim~(\ref{eq:Claim_Signature}) and the fact that for odd integers Chebyshev polynomials are odd functions (this is where the factor ${\rm sgn}(m)$ originates from). We have tested the analytic solution (\ref{eq:g(a)_Sin_Sin}) in a simple Monte Carlo study presented on Fig.~\ref{fig:Case_6_Sin_Sin_0a}, for $n=12$, $s_{12} = 0.25$ and $m=2,4,-4$.  
%


\section{Identities}
\label{app:Identities}


In this appendix we provide a detailed derivation of all non-trivial identities which were used in the main part of the paper. 

\subsection{Relation between Chebyshev polynomials and Bessel functions}
\label{app:Relation_between_Chebyshev_polynomials_and_Bessel_functions}

For the calculations of characteristic functions in Section~\ref{ss:Characteristic_functions}, we need to evaluate the following generic integral:
\begin{eqnarray}
\mathcal{I}_n &\equiv& \int_{-1}^{1}\frac{e^{ika}T_n(a)}{\sqrt{1-a^2}}\,da\,,
\label{eq:I_n}
\end{eqnarray}
where $T_n$ is Chebyshev polynomial of the first kind and $n$th order. We start from two table integrals (labeled as 7.355.1 and 7.355.2 in~\cite{GR:2007}):
\begin{eqnarray}
\int_0^1T_{2n+1}(a)\sin ka\frac{da}{\sqrt{1-a^2}} &=& (-1)^n\frac{\pi}{2}J_{2n+1}(k)\,, \qquad k>0\,,\\
\int_0^1T_{2n}(a)\cos ka\frac{da}{\sqrt{1-a^2}} &=& (-1)^n\frac{\pi}{2}J_{2n}(k)\,, \qquad k>0\,.
\end{eqnarray}
One can easily show straight from the definition of Chebyshev polynomial the following relation:
\begin{equation}
T_{n}(-x) = (-1)^nT_n(x)\,,
\label{eq:parity_Chebyshev}
\end{equation}
which determines the parity of Chebyshev polynomial of the first kind solely as a function of its order $n$. We can split $\mathcal{I}_n$ defined in (\ref{eq:I_n}) into two parts as:
\begin{equation}
\mathcal{I}_n = \int_{-1}^{1}\frac{\cos ka\,T_n(a)}{\sqrt{1-a^2}}\,da + i\,\int_{-1}^{1}\frac{\sin ka\,T_n(a)}{\sqrt{1-a^2}}\,da\,.
\label{eq:In_expanded}
\end{equation}
It is clear that due to result (\ref{eq:parity_Chebyshev}) the first integral in Eq.~(\ref{eq:In_expanded}) will be nonzero only for even $n$, while the second integral in Eq.~(\ref{eq:In_expanded}) will be nonzero only for odd $n$; we work out two possibilities separately.\\
%
\noindent{a) \textbf{$n$ even.}} We define $n\equiv 2l$, and it follows straightforwardly:
\begin{eqnarray}
\mathcal{I}_{2l} &=& \int_{-1}^{1}\frac{\cos ka\, T_{2l}(a)}{\sqrt{1-a^2}}\,da \nonumber\\
&=&2\times\int_{0}^{1}\frac{\cos ka\, T_{2l}(a)}{\sqrt{1-a^2}}\,da\nonumber\\
&=&(-1)^l\pi J_{2l}(k)\,.
\label{eq:I_2l}
\end{eqnarray}
\vspace{0.244cm}
\noindent{b) \textbf{$n$ odd.}} We define $n\equiv 2l+1$, and it follows analogously:
\begin{eqnarray}
\mathcal{I}_{2l+1} &=& i\int_{-1}^{1}\frac{\sin ka\, T_{2l+1}(a)}{\sqrt{1-a^2}}\,da \nonumber\\
&=&2i\times\int_{0}^{1}\frac{\sin ka\, T_{2l+1}(a)}{\sqrt{1-a^2}}\,da\nonumber\\
&=&i(-1)^l\pi J_{2l+1}(k)\,.
\label{eq:I_2l+1}
\end{eqnarray}
The above analytic results determine the relations between Chebyshev polynomials of the first kind and the Bessel function of the first kind. 


\subsection{Calculation of moments from characteristic functions}
\label{app:Calculation_of_moments_from_characteristic_functions.}

In this section we outline all technical details which led to the results presented in Section~\ref{s:Higher_order_moments}. We start with the well-known series representation of Bessel function of the first kind:
\begin{equation}
J_m(x)=\sum_{l=0}^\infty\frac{(-1)^l}{2^{2l+m}\,l!\,(l\!+\!m)!}\,x^{2l+m}\,,
\label{eq:Bessel_series}
\end{equation}
from which we can obtain the $n$th derivative of $m$th order Bessel function of the first kind evaluated at 0, i.e. $J^{(n)}_m(0)$. After some straightforward algebra we have obtained: 
\begin{equation}
J^{(n)}_m(0)=
\left\{
 \begin{array}{ll}
  \frac{(-1)^{\frac{n-m}{2}}\,n!}{2^n\left(\frac{n-m}{2}\right)!\left(\frac{n+m}{2}\right)!}\,,& n\!-\!m{\rm\ is\ an\ even\ nonnegative\ integer}\,, \\
  0\,,& {\rm otherwise}\,. \\  
 \end{array} 
\right.
\label{eq:Bessel_zeroes_nth_derivative}
\end{equation}
The above relation is the key result which we use solely to establish all the claims in subsequent derivations. 

We now calculate explicitly the $n$th algebraic moment of the real part of single-particle $Q$-vector evaluated in harmonic $m$, for the case when $n$ is an even integer. We write $n=2r$, where $r$ is some positive integer, and in the result (\ref{eq:mu'a_n_main_part}) we insert (\ref{eq:CharacteristicFunction_fullFS_cosmx}) to obtain:
\begin{equation}
\mu'_{{\rm Re}\,u_m,n}=i^{-2r}\bigg[J_{0}^{(2r)}(0)+2\sum_{p=1}^\infty\,(-1)^p\left[c_{2p\cdot m}J_{2p}^{(2r)}(0)
-  i c_{(2p-1)\cdot m} J_{2p-1}^{(2r)}(0)\right]\bigg]\,, \qquad n=2r\,.
\label{eq:temp_20141004}
\end{equation}
Solely from the result (\ref{eq:Bessel_zeroes_nth_derivative}) we can establish the following two claims: a) The second term within summation in (\ref{eq:temp_20141004}) yields no contribution due to the fact that $2r-(2p-1)$ is always an odd integer; b) There is a {\it finite} number of terms in the first term within summation in (\ref{eq:temp_20141004}) which can yield to nonzero contributions due to the fact that $2r-2p$ is an even nonnegative integer only for $p\leq r$. It follows:
\begin{eqnarray}
\mu'_{{\rm Re}\,u_m,n}&=&(-1)^r\bigg[\frac{(-1)^r(2r)!}{4^r(r!)^2}+2\sum_{p=1}^r\,(-1)^pc_{2p\cdot m}\frac{(-1)^{r-p}(2r)!}{4^r(r\!-\!p)!(r\!+\!p)!}\bigg]\nonumber\\
 &=&
  \frac{(2r)!}{4^r(r!)^2}\bigg[1\!+\!2\sum_{p=1}^r c_{2p\cdot m}\frac{(r!)^2}{(r\!-\!p)!(r\!+\!p)!}\bigg]\,, \qquad n=2r\,,
\end{eqnarray}
which is the first part of result (\ref{eq:nth_moment_of_real_part_of_single-particle_Q-vector_evaluated_in_harmonic_m}) presented in the main part which applies for the cases when $n$ is an even integer. In a completely analogous way we have derived the other three cases presented in Eqs.~(\ref{eq:nth_moment_of_real_part_of_single-particle_Q-vector_evaluated_in_harmonic_m}) and (\ref{eq:nth_moment_of_imaginary_part_of_single-particle_Q-vector_evaluated_in_harmonic_m}) in the main part. 


\section{Moments}
\label{app:Moments}


In this self-contained appendix we summarize the procedure which can be used in order to determine the moments of a function $a(x)$ of a random observable $x$ even without the explicit knowledge of a p.d.f. $g(a)$ of a function $a(x)$. The procedure presented here is fairly general and applies also for the cases of multivariate functions $a(x_1,x_2,\ldots)$ which are of our main interest. The material in this appendix is heavily based on the basic material presented in the classical textbook~\cite{Cowan:1998ji} and is further generalized and expanded. Throughout this appendix the expectation value of any random observable $x$ is indicated by $E[x]$.


\subsection{Moments of a random observables}
\label{app:Moments_of_a_random_observables}

If the starting random observable is denoted by $x$, its sample space by $X$ and its p.d.f. by $f(x)$ then the $n$th algebraic moment of $x$ we denote by $\mu_n'$ and define as~\cite{Cowan:1998ji}:
\begin{equation}
\mu'_n = E[x^n] = \int_X x^n f(x)\,dx\,.
\label{eq:mu'_n}
\end{equation}
From (\ref{eq:mu'_n}) we see that the mean, $\mu$, is equal to the first algebraic moment $\mu'_{1}$. On the other hand, the $n$th central moment is denoted as $\mu_n$ and defined as~\cite{Cowan:1998ji}:
\begin{equation}
\mu_n = E[(x-E[x])^n] = \int_X(x-\mu)^n f(x)\,dx\,.
\label{eq:mu_n}
\end{equation}
In what follows we will calculate directly only the algebraic moments (\ref{eq:mu'_n}), and use the following well-known relation to obtain the central moments:
\begin{equation}
\mu_n =\sum_{k=0}^{n}\binom{n}{k}(-1)^{n-k}\mu_k'\,\mu^{n-k},\qquad \mu_0'=1\,.
\label{eq:relationBetweenAlgebraicAndCentralMoments}
\end{equation}
The above relation yields for the lower order moments the following explicit results:
\begin{eqnarray}
\mu_1 &=& 0\,,\\ 
\mu_2 &=& -\mu^2+\mu_2'\,,\\
\mu_3 &=& 2\mu^3-3\mu\mu_2'+\mu_3'\,,\\
\mu_4 &=& -3\mu^4+6\mu^2\mu_2'-4\mu\mu_3'+\mu_4'\,.
\label{eq:lowerOrderCases}
\end{eqnarray}
Instead of reporting the third and the fourth moment, it is more customary to report skewness ($\gamma_1$) and kurtosis ($\gamma_2$), for which we use the following conventions:
\begin{eqnarray}
\gamma_1 &\equiv&\frac{\mu_3}{\mu_2^{3/2}}\,,\label{eq:skewness}\\
\gamma_2 &\equiv&\frac{\mu_4}{\mu_2^2}-3\,.\label{eq:kurtosis} 
\end{eqnarray}
With above conventions, both skewness and kurtosis are zero for Gaussian p.d.f. On the other hand, the function of $x$ we denote by $a(x)$, its sample space by $A$ and its p.d.f. by $g(a)$. Then the $n$th algebraic moment of $a$ is given by~\cite{Cowan:1998ji}:
\begin{eqnarray}
E[a^n] &=& \int_A a^n g(a) da\nonumber\\
&=&\int_X [a(x)]^n f(x) dx\nonumber\\
&\equiv&\mu'_{a,n}\,.
\label{eq:mu'_an}
\end{eqnarray}
The above relation indicates that even without the explicit knowledge of $g(a)$ we can still obtain the  moments of $a$ only from the knowledge of p.d.f. $f(x)$ and the functional dependence $a(x)$. 

The result (\ref{eq:mu'_an}) easily generalizes to the case of multivariate function of random observables.
In this context, we utilize the different notation for the starting random observable in order to come closer to their physical interpretation in our subsequent calculations. For a set of $M$ random observables $(\varphi_1, \ldots, \varphi_M)$ with sample spaces $(\Phi_1, \ldots, \Phi_M)$ on which multivariate function $a$ depends we introduce the shortcut notation $\underline{\varphi} \equiv (\varphi_1, \ldots, \varphi_M)$. Then we have for the $n$th algebraic moment of multivariate function $a$~\cite{Cowan:1998ji}:
\begin{eqnarray}
E[a^n] &=& \int_A a^n g(a)\,da\nonumber\\
&=&\int_{\Phi_1}\cdots\int_{\Phi_M}[a(\underline{\varphi})]^n f(\underline{\varphi})\,d\varphi_1\cdots d\varphi_M\nonumber\\
&\equiv&\mu'_{a,n}\,,
\label{eq:mu_a_multivariate}
\end{eqnarray}
where $g(a)$ is p.d.f. of $a$, while $f(\underline{\varphi})$ is joint multivariate p.d.f. of random observables $(\varphi_1, \ldots, \varphi_M)$. 

The above very general results we now apply for the specific context of anisotropic flow analysis. For the case when only anisotropic flow correlations are present, we have the following factorization of multivariate p.d.f.:
\begin{equation}
f(\underline{\varphi}) = f_{\varphi_{1}}(\varphi_{1})\cdots f_{\varphi_{M}}(\varphi_M)\,,
\label{eq:factorization_in_appendix}
\end{equation}
where for each $i$ the functional form of normalized marginalized p.d.f. $f_{\varphi_i}(\varphi_i)$ is the same and it equals to the following Fourier-like p.d.f.:
\begin{equation}
f_{\varphi_i}(\varphi_i)=\frac{1}{2\pi}\bigg[1+2\sum_{n=1}^\infty (c_n\cos n\varphi_i + s_n\sin n\varphi_i)\bigg]\,,\qquad\forall i\,.
\label{eq:Fourier_cn_sn_Appendix}
\end{equation}
or equivalently using an alternative parametrization: 
\begin{equation}
f_{\varphi_i}(\varphi_i) = \frac{1}{2\pi}\bigg[1+2\sum_{n=1}v_n\cos[n(\varphi_i-\Psi_n)]\bigg]\,,\qquad\forall i\,.
\label{eq:Fourier_marginalized}
\end{equation}
The two assumptions, namely the factorization of multivariate p.d.f. in (\ref{eq:factorization_in_appendix}) and the equality of all single-particle p.d.f.'s $f_{\varphi_i}(\varphi_i)$ in (\ref{eq:Fourier_cn_sn_Appendix}) or (\ref{eq:Fourier_marginalized}) tremendously simplify the evaluation of moments defined in (\ref{eq:mu_a_multivariate}). The major simplification comes from the fact that a lot of single-particle integrals in Eq.~(\ref{eq:mu_a_multivariate}) are exactly the same apart from trivial {\it relabeling} of integration variables. Based on this conclusion, we demonstrate in the next section that in practice it suffices to perform integration only for few distinct generic terms, and all other contributions can be accounted for with combinatorial coefficients. 


\subsection{Generic integration}
\label{app:Generic_integration}

We start by observing that the term $[a(\underline{\varphi})]^n$ in (\ref{eq:mu_a_multivariate}) can be decomposed in general as follows by using the multinomial theorem:
\begin{equation}
[a(\underline{\varphi})]^n\equiv (a_1+a_2+\cdots+a_M)^n = \sum_{k_1+k_2+\cdots+k_M=n}\frac{n!}{k_1!k_2!\cdots k_M!}\, a_1^{k_1}a_2^{k_2}\cdots a_M^{k_M}\,.
\label{eq:multinomial_theorem}
\end{equation}
We will work out now the above expression explicitly for arbitrary $M$ and for few practical cases of interest for $n$. We first conclude that all generic terms in integration (\ref{eq:mu_a_multivariate}) are generated with the special case $M=n$, and for $M>n$ all those generic terms reappear scaled with factors which depends only on $M$ that we now determine. Due to relabeling, those scaling factors have two separate contributions; the first one from multinomial coefficient which can be obtain straightforwardly from (\ref{eq:multinomial_theorem}), and the second one which can be determined from the {\it weak composition} of natural number $n$ which do not trivially differ by permutations. Weak composition of natural number $n$ is by definition a number of ways of writing $n$ as the sum of a sequence of non-negative integers. We now illustrate this procedure by working out the concrete cases of interest. 

\subsubsection{$n$=2}
\label{app:n=2}

\noindent The expression which will determine all generic terms in integration is simply:
\begin{equation}
(a_1+a_2)^2\,.
\end{equation}
The weak composition of $n=2$ is the following sequence:
\begin{equation}
\{0,2\}, \{1,1\}, \{2,0\}\,,
\end{equation}
from which we select the terms which do not trivially differ by permutations to obtain:
\begin{equation}
\{0,2\}, \{1,1\}\,.
\end{equation}
We can now identify the corresponding generic terms in integration as:
\begin{eqnarray}
\{0,2\} &\mapsto& a_1^2\,,\nonumber\\
\{1,1\} &\mapsto& a_1a_2\,,
\end{eqnarray}
while their multinomial coefficients are 
\begin{eqnarray}
\{0,2\} &\mapsto& \frac{2!}{0!2!}=1\,,\nonumber\\
\{1,1\} &\mapsto& \frac{2!}{1!1!}=2\,.
\end{eqnarray}
For $M>2$ the above results generalize into the following sets, from which we can trivially determine the combinatorial coefficients:
\begin{eqnarray}
\{0,0,\ldots,0,2\} &\mapsto& \frac{M!}{(M-1)!}\,,\nonumber\\
\{0,0,\ldots,0,1,1\} &\mapsto& \frac{M!}{(M-2)!2!}\,.
\end{eqnarray}
Putting up everything together, we have obtained finally the following generic expression:
\begin{equation}
(a_1+a_2+\ldots+a_M)^2\propto a_1^2\times 1\times\frac{M!}{(M-1)!} + a_1a_2 \times 2 \times \frac{M!}{(M-2)!2!}\,.
\label{eq:genericIntegration_n=2}
\end{equation}
The above result indicates that in calculating the 2nd algebraic moment of a multivariate function with (\ref{eq:mu_a_multivariate}) we only need to perform two generic integrals. For instance, we can use the above results to calculate the moments of the real and imaginary parts of $M$-particle $Q$-vectors analytically as follows. With the definitions 
\begin{eqnarray}
a(\underline{\varphi})&\equiv& {\rm Re}\,Q_n = \sum_{i=1}^{M}\cos n\varphi_i\,,\label{eq:aReQ}\\ 
a(\underline{\varphi})&\equiv& {\rm Im}\,Q_n = \sum_{i=1}^{M}\sin n\varphi_i\,,\label{eq:aImQ}
\end{eqnarray}
from (\ref{eq:genericIntegration_n=2}) we have immediately:
\begin{eqnarray}
E[({\rm Re}\,Q_n)^2] &=& 1 \times \frac{M!}{(M-1)!}\int_{0}^{2\pi} \cos^2 n\varphi_1f(\varphi_1) \,d\varphi_1\\
             &&{}+ 2 \times \frac{M!}{(M-2)!2!}\int_{0}^{2\pi}\int_{0}^{2\pi} \cos n\varphi_1\cos n\phi_2f(\varphi_1)f(\varphi_2)\, d\varphi_1 d\varphi_2\nonumber\,. 
\end{eqnarray}
In order to obtain the above relation we have used the factorization property (\ref{eq:factorization_in_appendix}) of joint multivariate p.d.f. and the fact that single particle p.d.f.'s in (\ref{eq:Fourier_marginalized}) are normalized to unity. Using the orthogonality properties of trigonometric functions~(\ref{eq:sin_sin}-\ref{eq:sin_cos}), both integrals above can be solved analytically for the most general case of single-particle Fourier-like p.d.f. in (\ref{eq:Fourier_marginalized}), parametrized with all flow harmonics and symmetry planes. We have obtained for the most general case: 
\begin{eqnarray}
E[({\rm Re}\,Q_n)^2] &=& M \frac{1}{2} \left(1\!+\!v_{2n}\cos 2n \Psi_{2n}\right)\!+\!M(M\!-\!1)(v_n\cos n \Psi_{n})^2\,,\\
E[({\rm Im}\,Q_n)^2] &=& M \frac{1}{2} \left(1\!-\!v_{2n}\cos 2n \Psi_{2n}\right)\!+\!M(M\!-\!1)(v_n\sin n \Psi_{n})^2\,,\\
E[|Q_n|^2] &=& E[({\rm Re}\,Q_n)^2]\!+\!E[({\rm Im}\,Q_n)^2]\nonumber\\
           &=& M[v_n^2(M\!-\!1)\!+\!1]\,.
\end{eqnarray}
We remark that we have used the generic integration in (\ref{eq:genericIntegration_n=2}) for two specific choices of $a_i$, namely $a_i\equiv\cos n\varphi_i$ in (\ref{eq:aReQ}) and $a_i\equiv\sin n\varphi_i$ in (\ref{eq:aImQ}), but this procedure is much more general. In particular, for $a_i$ we can in general case select any univariate function of $\varphi_i$, e.g. $a_i\equiv\cos n\varphi_i+\sin n\varphi_i$, etc. We now briefly summarize the analogous procedure for the cases $n=3$ and $n=4$.

\subsubsection{n=3}
\label{app:n=3}

\noindent The expression which will determine all generic terms in the integration (\ref{eq:mu'_an}) is:
\begin{equation}
(a_1+a_2+a_3)^3\,.
\end{equation}
The weak composition of $n=3$ is the following sequence:
\begin{equation}
\{0, 0, 3\}, \{0, 1, 2\}, \{0, 2, 1\}, \{0, 3, 0\}, \{1, 0, 2\}, \{1, 1, 1\}, \{1, 2, 0\}, \{2, 0, 1\}, \{2, 1, 0\}, \{3, 0, 0\}\,,
\end{equation}
from which we select the terms which do not trivially differ by permutations to obtain:
\begin{equation}
\{0, 0, 3\}, \{0, 1, 2\}, \{1, 1, 1\}\,.
\end{equation}
We can now identify the corresponding generic terms in the integration in (\ref{eq:mu'_an}) as:
\begin{eqnarray}
\{0, 0, 3\} &\mapsto& a_1^3\,,\nonumber\\
\{0, 1, 2\} &\mapsto& a_1a_2^2\,,\nonumber\\
\{1, 1, 1\} &\mapsto& a_1a_2a_3\,,
\end{eqnarray}
with multinomial coefficients 
\begin{eqnarray}
\{0, 0, 3\} &\mapsto& \frac{3!}{0!0!3!}=1\,,\nonumber\\
\{0, 1, 2\} &\mapsto& \frac{3!}{0!1!2!}=3\,,\nonumber\\
\{1, 1, 1\} &\mapsto& \frac{3!}{1!1!1!}=6\,.
\end{eqnarray}
For $M>3$ we obtain the following sets from which we calculate the combinatorial coefficients:
\begin{eqnarray}
\{0,0,\ldots,0,3\} &\mapsto& \frac{M!}{(M-1)!}\,,\nonumber\\
\{0,0,\ldots,0,1,2\} &\mapsto& \frac{M!}{(M-2)!}\,,\nonumber\\
\{0,0,\ldots,0,1,1,1\} &\mapsto& \frac{M!}{(M-3)!3!}\,.
\end{eqnarray}
Putting up everything together, we have obtained our final expression for the generic integration in (\ref{eq:mu'_an}) for arbitrary $M$ and $n=3$:
\begin{eqnarray}
(a_1+a_2+\ldots+a_M)^3&\propto& a_1^3\times 1\times\frac{M!}{(M-1)!}\nonumber \\
&+& a_1a_2^2 \times 3 \times \frac{M!}{(M-2)!}\nonumber\\
&+& a_1a_2a_3 \times 6 \times \frac{M!}{(M-3)!3!}\,.
\label{eq:genericIntegration_n=3}
\end{eqnarray}

\subsubsection{n=4}
\label{app:n=4}

\noindent By following the completely analogous procedure as for the previous cases $n=2$ and $n=3$, we have obtained the following expression for the generic integration in (\ref{eq:mu'_an}) which is valid for arbitrary $M$ and $n=4$:
\begin{eqnarray}
(a_1+a_2+\ldots+a_M)^4&\propto& a_1^4\times 1\times\frac{M!}{(M-1)!}\nonumber \\
&+& a_1a_2^3 \times 4 \times \frac{M!}{(M-2)!}\nonumber\\
&+& a_1^2a_2^2 \times 6 \times \frac{M!}{(M-2)!2!}\nonumber\\
&+& a_1a_2a_3^2 \times 12 \times \frac{M!}{(M-3)!2!}\nonumber\\
&+& a_1a_2a_3a_4 \times 24 \times \frac{M!}{(M-4)!4!}\,.
\label{eq:genericIntegration_n=4}
\end{eqnarray}
We now use this technology and calculate exactly for the most general case of multichromatic flow the higher order moments of few observables of interest. 


\subsection{Exact results for the moments}
\label{app:Exact_results_for_the_moments}

For the even moments of $Q$-vector amplitudes for the most general case of multichromatic flow parametrized as in Eq.~(\ref{eq:Fourier_cn_sn_Appendix}) we have obtained the following exact results:
\begin{equation}
E[|Q_m|^2] = M\big[1\!+\!(M\!-\!1)(c_m^2\!+\!s_m^2)\big]\,,
\label{eq:E|Qm|^2}
\end{equation}
\begin{eqnarray}
E[|Q_m|^4] &=& M\big[2M\!-\!1\!+\!(M\!-\!3)(M\!-\!2)(M\!-\!1)(c_m^2\!+\!s_m^2)^2\!+\!4(M\!-\!1)^2(c_m^2\!+\!s_m^2)\nonumber\\
&&{}+\!(M\!-\!1)(c_{2m}^2\!+\!s_{2m}^2)\!+\!2(M\!-\!2)(M\!-\!1)c_{2m}(c_m^2\!-\!s_m^2)\!+\!4(M\!-\!2)(M\!-\!1)c_{m} s_m s_{2m}\big]\,,
\label{eq:E|Qm|^4}
\end{eqnarray}
\begin{eqnarray}
E[|Q_m|^6] &=& M\big[6M^2\!-\!9M\!+\!4\!+\!(M\!-\!5)(M\!-\!4)(M\!-\!3)(M\!-\!2)(M\!-\!1)(c_m^2\!+\!s_m^2)^3\nonumber\\
&&{}\!+\!9(M\!-\!3)(M\!-\!2)^2(M\!-\!1)(c_m^2\!+\!s_m^2)^2\!+\!3(6M^2\!-\!15M\!+\!11)(M\!-\!1)(c_m^2\!+\!s_m^2)\nonumber\\
&&{}\!+\!3(3M\!-\!4)(M\!-\!1)(c_{2m}^2\!+\!s_{2m}^2)\!+\!(M\!-\!1)(c_{3m}^2\!+\!s_{3m}^2)\nonumber\\
&&{}\!+\!6(M\!-\!4)(M\!-\!3)(M\!-\!2)(M\!-\!1)c_{2m}(c_m^4\!-\!s_m^4)\nonumber\\
&&{}\!+\!9(M\!-\!3)(M\!-\!2)(M\!-\!1)(c_m^2\!+\!s_m^2)(c_{2m}^2\!+\!s_{2m}^2)\nonumber\\
&&{}\!+\!6(3M\!-\!5)(M\!-\!2)(M\!-\!1)(c_m^2c_{2m}\!+\!2c_m s_m s_{2m}\!-\!s_m^2c_{2m})\nonumber\\
&&{}\!+\!12(M\!-\!4)(M\!-\!3)(M\!-\!2)(M\!-\!1)s_{2m}(c_m^3s_m\!+\!s_m^3c_m)\nonumber\\
&&{}\!+\!2(M\!-\!3)(M\!-\!2)(M\!-\!1)(c_m^3c_{3m}\!-\!s_m^3s_{3m})\nonumber\\
&&{}\!+\!6(M\!-\!3)(M\!-\!2)(M\!-\!1)(c_m^2s_ms_{3m}\!-\!s_m^2c_mc_{3m})\nonumber\\
&&{}\!+\!6(M\!-\!2)(M\!-\!1)(c_mc_{2m}c_{3m}\!-\!s_ms_{2m}c_{3m}\!+\!s_mc_{2m}s_{3m}\!+\!c_ms_{2m}s_{3m})\big]\,.
\label{eq:E|Qm|^6}
\end{eqnarray}
From the above moments one can obtain straightforwardly the exact and most general results for the variance and skewness of two-particle correlations for the most general case of multichromatic flow. For instance, from the Eq.~(\ref{eq:2pBasicExample_first_part}) it follows trivially that:
\begin{eqnarray}
\mu'_{\left<2\right>,2} &=& \frac{E[(|Q_m|^2\!-\!M)^2]}{M^2(M\!-\!1)^2}\nonumber\\
&=&\frac{E[|Q_m|^4]\!-\!2ME[|Q_m|^2]\!+\!M^2}{M^2(M\!-\!1)^2}\,,
\label{eq:mu'_2p_2}
\end{eqnarray}
so that after some algebra, taking into account the result (\ref{eq:2pBasicExample_second_part}), we have the following exact result for the variance $\sigma_{\left<2\right>}^2$:
\begin{eqnarray}
\sigma_{\left<2\right>}^2 &\equiv& \mu'_{\left<2\right>,2} - \mu_{\left<2\right>}^2\nonumber\\
&=& \frac{1}{M(M\!-\!1)}\big[1\!-\!2(2M\!-\!3)(c_m^2\!+\!s_m^2)^2\!+\!2(M\!-\!2)(c_m^2\!+\!s_m^2)\nonumber\\
&&{}\!+\!c_{2m}^2\!+\!s_{2m}^2\!+\!2(M\!-\!2)c_{2m}(c_m^2\!-\!s_m^2)\!+\!4(M\!-\!2)c_ms_ms_{2m}\big]\,.
\label{eq:variance_Appendix}
\end{eqnarray}
This is the generalization of result presented recently as Eq.~(6) in~\cite{Bilandzic:2013kga}, which was obtained under the assumption that the initial single-particle Fourier-like p.d.f.~(\ref{eq:Fourier_marginalized}) was parametrized only with the amplitudes $v_n$. In a completely analogous way one can also obtain the exact result for the skewness~(\ref{eq:skewness}) of two-particle azimuthal correlation only from the knowledge of the above even moments of $Q$-vector amplitudes. One start with the calculation of third algebraic moments as:
\begin{eqnarray}
\mu'_{\left<2\right>,3} &=& \frac{E[(|Q_m|^2\!-\!M)^3]}{M^3(M\!-\!1)^3}\nonumber\\
&=&\frac{E[|Q_m|^6]\!-\!3ME[|Q_m|^4]\!+3M^2E[|Q_m|^2]-\!M^3}{M^3(M\!-\!1)^3}\,,
\label{eq:mu'_2p_3}
\end{eqnarray}
and the rest of calculation follows immediately.  


\end{document}